\documentclass{aastex631}
\usepackage{longtable}

\accepted{April 11, 2024}

\begin{document}

\title{The Timescales of Star Cluster Emergence: The Case of NGC\,4449}

\author[0009-0001-5214-6330]{Timothy McQuaid}
\affiliation{Department of Astronomy, University of Massachusetts Amherst, 710 North Pleasant Street, Amherst, MA 01003, USA}
\affiliation{Department of Astronomy, New Mexico State University, Las Cruces, NM 88003, USA}

\author[0000-0002-5189-8004]{Daniela Calzetti}
\affiliation{Department of Astronomy, University of Massachusetts Amherst, 710 North Pleasant Street, Amherst, MA 01003, USA}

\author[0000-0002-1000-6081]{Sean T. Linden}
\affiliation{Department of Astronomy, University of Massachusetts Amherst, 710 North Pleasant Street, Amherst, MA 01003, USA}

\author[0000-0003-1427-2456]{Matteo Messa}
\affiliation{INAF – OAS, Osservatorio di Astrofisica e Scienza dello Spazio di Bologna, via Gobetti 93/3, I-40129 Bologna, Italy}
%\affiliation{Department of Astronomy, Universit\'e de Gen\`eve, 24 rue du G\'en\'eral-Dufour, 1211 Gen\`eve 4, Switzerland }
%\affiliation{Department of Astronomy, Stockholm University, Stockholm, Sweden}

\author[0000-0002-8192-8091]{Angela Adamo}
\affiliation{Department of Astronomy, Stockholm University, Stockholm, Sweden}

% \author[0000-0002-1723-6330]{Bruce Elmegreen}
% \affiliation{IBM Research Division, T.J. Watson Research Center, Yorktown Heights, NY, USA}

\author[0000-0002-1723-6330]{Bruce Elmegreen}
\affiliation{Katonah, NY 10536 USA}

\author[0000-0002-3247-5321]{Kathryn Grasha}
\affiliation{Research School of Astronomy and Astrophysics, Australian National University, Camberra, Australia}

\author[0000-0001-8348-2671]{Kelsey E. Johnson}
\affiliation{Department of Astronomy, University of Virginia, Charlottesville, VA, USA}

\author[0000-0002-0806-168X]{Linda J. Smith}
\affiliation{Space Telescope Science Institute, 3700 San Martin Drive, Baltimore, MD 21218, USA}

\author{Varun Bajaj}
\affiliation{Space Telescope Science Institute, 3700 San Martin Drive, Baltimore, MD 21218, USA}

%\author{et al. from LEGUS--IR Team}

\begin{abstract}

We survey the young star cluster population in the dwarf galaxy NGC\,4449 with the goal of investigating how stellar feedback may depend on the clusters' properties. Using Ultraviolet(UV)-optical-NearIR(NIR) photometry obtained from the \emph{Hubble Space Telescope}, we have recovered 99 compact sources exhibiting emission in the Pa$\beta$ hydrogen recombination line. Our analysis reveals these sources possess masses $10^{2}<M_{\odot}<10^{5}$, ages 1--20 Myr, and color excess E(B - V) in the range 0--1.4. After selecting clusters with mass above 3,000M$_{\odot}$ to mitigate stochastic sampling of the stellar initial mass function, we find that our IR--selected clusters have a median mass of $\sim$7$\times{10^{3}\text{ M}_{\odot}}$ and remain embedded in their surrounding gas and dust for 5--6 Myr. In contrast, line-emitting sources selected from existing UV/optically catalogs have a median mass of $\sim$3.5$\times{10^{4}\text{ M}_{\odot}}$ and have cleared their surroundings by 4 Myr. We further find that the environment in NGC\,4449 is too low pressure to drive these differences. We interpret these findings as evidence that the clearing timescale from pre--supernova and supernova feedback is cluster mass--dependent. Even in clusters with mass$\sim$7,000~M$_{\odot}$, stochastic sampling of the upper end of the stellar initial mass function is present, randomly decreasing the number of massive stars available to inject energy and momentum into the surrounding medium. This effect may increase the clearing timescales in these clusters by decreasing the effectiveness of both pre--supernova and supernova feedback; neither models nor observations have so far explored such dependence explicitly. Future studies and observations with, e.g., the JWST, will fill this gap.

\end{abstract}



\section{Introduction} \label{sec:intro}

\setcounter{page}{1}

Star formation occurs predominantly in clusters that emerge from the gravitational collapse of massive clouds of molecular gas and dust. When a molecular cloud cools sufficiently, its gravitational potential energy overcomes the kinetic energy of the gas, forming overdensities which then feed the formation of new stars \citep{Krumholz2019}. The formation of massive stars is however complicated by the presence of numerous counteracting mechanisms. While molecular gas, under the force of gravity, accretes onto a forming star, the star simultaneously injects energy into its surroundings \citep{Pellegrini+2011, Dale+2012, Krause+2013}. This energy, referred to as stellar feedback, comes in several forms including stellar winds, photoionization, radiation pressure and supernovae explosions. Feedback is able to counteract the force of gravity and expel the gas and dust surrounding the region of star formation \citep{Lucas+2020, Grudic+2022}. When enough material has been cleared, star formation effectively ceases within the cluster.
Furthermore, the energy and momentum injected into the surrounding interstellar  medium  drives turbulence and limits the amount of cold, dense gas that can collapse, thus contributing to the low star formation efficiency\citep{Hennebelle2011, Hopkins+2012, Dobbs+2015, Goldbaum+2016}. 
The impact of feedback on star formation is so significant that the conversion efficiency of turning molecular gas into stars is on the order of a few \% \citep{Ostriker+2010, Hopkins+2014, Hopkins+2020, Grudic+2018}. 
%This means that a significant amount of molecular gas is expelled from the cluster before it can be utilized for star formation. 
Therefore, stellar feedback is not only critical to regulating star formation on a local level, but on a galactic level as well \citep{Krumholz+2005, Ceverino2009, Dobbs+2011, Krumholz2019}.  

%High mass stars are responsible for contributing the vast majority of feedback in a galaxy. Since a star's luminosity scales with a steep power law with respect to mass $\sim L\propto M^{3}$ , high mass stars consequently produce a far greater amount of high energy photons than lower mass stars; thus they are responsible for injecting the majority of energy into the surrounding medium. However, the lifespans of the highest mass stars, O-type and B-type, are incredibly short on a cosmic timescale, only 3-10 Myr \citep{Leitherer+2014}. Furthermore, the demise of these high mass stars can result in a supernovae, which releases even more energy into the surrounding medium. There is a high degree of uncertainty as to whether pre-supernova feedback can sufficiently clear away the gas within the cluster before the occurrence of supernovae. If the clearing time is indeed shorter than the timescale of supernovae, then the supernovae can inject a greater amount of metals and energy into the ISM.
 
Massive star feedback is conventionally divided into pre--supernova (pre--SN) feedback (photoionization, direct and indirect radiation pressure and stellar winds) and supernova feedback. Pre--SN acts on short timescales, typically $\lesssim$4~Myr \citep{Pellegrini+2011, Dale+2012, Krause+2013, Krumholz2019}, and helps clear the medium before supernova explosions occur starting around 4~Myr \citep{Leitherer+2014}. Some models indicate that radiative feedback may be key for regulating star formation within galaxies \citep{Hopkins+2020, Bending+2022}. 

A number of prior observational studies have concluded that pre-SN feedback is the main mechanism for regulating star formation in clusters; these conclusions are drawn from measurements of the timescales necessary for the clusters to emerge from their natal clouds and become optically visible. Several of these studies have examined UV/optical photometry from the \emph{Hubble Space Telescope} (HST) of young star clusters to derive clearing timescales $<$4-5 Myr and as short as 2 Myr, in nearby galaxies \citep{Whitmore+2011,Hollyhead+2015, Hannon+2019, Hannon+2022}. However, these surveys, which only utilize optical and UV observations, are inherently limited by the effects of dust attenuation, and may miss highly dust-attenuated clusters, i.e., the clusters most closely associated with their natal clouds. Several other studies have attempted to remedy this by including CO data from ground--based observatories or 24 $\mu$m observations from the \emph{Spitzer Space Telescope} \citep{Matthews+2018,Grasha+2018,Grasha+2019,Kruijssen+2019,Kim+2021,Chevance+2022}. These studies have similarly derived a short clearing time, 3-5 Myr. The use of CO data or infrared images is, however, limited by low resolution. For instance, CO data typically subtends ~50 pc, far greater than the ~3 pc radius of young clusters \citep{Ryon+2017, Brown+2021}. Although \citet{Corbelli+2017} leverages the proximity of the Local Group galaxy M\,33 (distance $\sim$840~kpc) to increase the spatial resolution of the {\em Spitzer  Space Telescope}'s 24~$\mu$m images and conclude that the embedded phase of star  formation is short,  $\sim$2~Myr, the subtended spatial scale is still 26~pc, much larger than a single star cluster.

\citet{Messa2021} utilized HST near-UV to near-IR photometry to study the cluster populations of the nearby galaxy NGC\,1313. They found between 40-60$\%$ of young clusters ($<$6 Myr) are missed when selection is only based on UV and optical data. Their analysis also extended the clearing time from ~2-3 Myr to ~3-4~Myr. \citet{Messa2021} is however restricted by the requirement that sources still need to be detectable at optical wavelengths in order to determine their ages and masses. \citet{Calzetti2023}, using HST optical and IR images of NGC\,4449, attempted to obtain a sample of highly extincted clusters by selecting sources that emit in Pa$\beta$ but are undetected at optical and shorter wavelengths. The specific selection requirements used in that study resulted in a limited number of sources (34) with a narrow range of properties, which leaves open the question of whether those authors' approach includes potential biases. 

The studies above, with their variety of timescales,  hint at the possibility that the clearing timescale may be influenced by the galactic environment and/or by the cluster's properties. The dependence of pre--SN feedback on galactic environment has been investigated by \citet{McLeod+2021}, \citet{DellaBruna+2022} and \citet{Chevance+2022} for a few local spiral galaxies. These authors find that the galactic environment does have an influence on the effectiveness of pre--SN feedback. While \citet{Chevance+2022} conclude that presence of morphological features introduce variations in the pre--SN clearing timescales, \citet{McLeod+2021} find a dependency on galactocentric distance. \citet{McLeod+2021} conclude that supernova feedback is enhanced by the clearing done by pre--SN feedback in the outer, less dense regions of galaxies. Similarly, \citet{DellaBruna+2022} find that the HII regions in the center of the starburst galaxy M83  are confined by the ambient pressure, while they are over--pressurized and expanding in the disk.  

In this work we extend the analysis of \citet{Messa2021} to the nearby Magellanic Irregular dwarf galaxy NGC\,4449. NGC\,4449 is located at a distance of 4.2 Mpc \citep{Tully+2013}, implying that our HST nearIR imaging data resolves regions as small as $\lesssim$2--3~pc, i.e., comparable to a cluster's size \citep{Ryon+2017, Brown+2021}. With a star formation rate of approximately 0.5 M$_\odot\ $ yr$^{-1}$ and a stellar mass of $\sim$10$^9$~M$_{\odot}$ , NGC\,4449 is located about 3$\sigma$ above the Star Formation Main Sequence for local galaxies \citep{Cook+2014} which qualifies it as a starburst. The metallicity of the galaxy is about 40\% solar\footnote{We adopt a value 12+log(O/H)=8.69 for the solar oxygen abundance \citep{Asplund+2009}.} and has a modest galactocentric gradient \citep{Berg+2012, Pilyugin+2015}. There are several similarities between NGC\,4449 and NGC\,1313 \citep[the latter studied by][]{Messa2021}. The two galaxies have comparable distances, oxygen abundances and specific SFRs (SFR/M$_*$) \citep[see,][]{Messa2021}. Within a factor $\sim$3, their masses are comparable to that of the Large Magellanic Cloud. However, they differ in their SFR surface density, with NGC\,4449 having $\Sigma_{SFR}=0.03$~M$_{\odot}$~yr$^{-1}$~kpc$^{-2}$, or three times the SFR surface density of NGC\,1313 in the regions targeted by HST \citep{Messa2021, Calzetti2023}. 

\begin{figure}
    
    \centering
    \includegraphics[width = 0.8\linewidth]{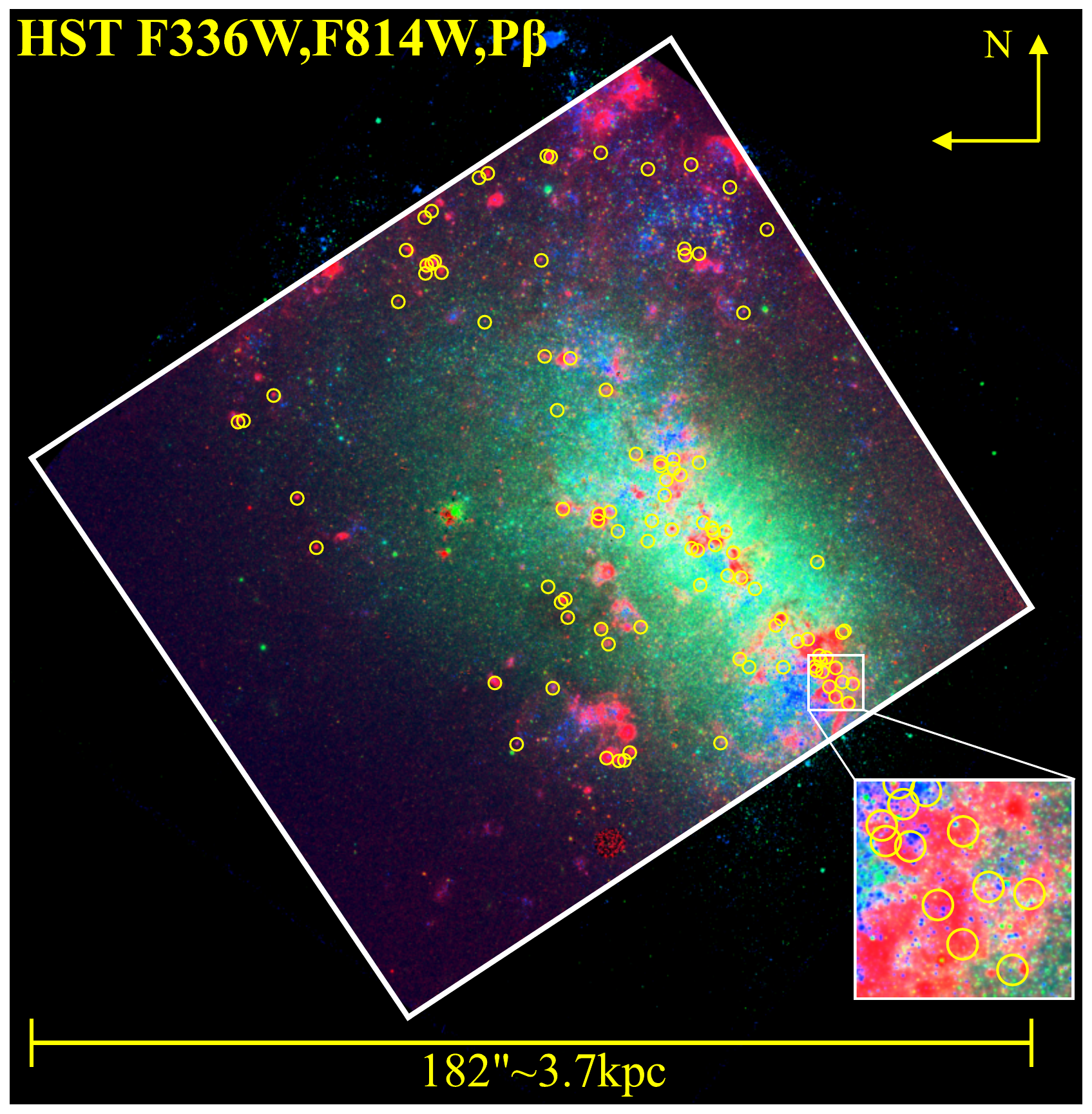}
    \caption{A three color composite of NGC\,4449 (blue=F336W, green=F814W, red=Pa$\beta$ emission). The locations of the 99 cluster candidates are marked by yellow circles. Due to the requirement that sources be detected in Pa$\beta$, all sources lie within the field of view of the F128N filter, traced by the white line. North is up, East is left.}
    \label{fig:Field}
\end{figure}

We leverage the same HST UV--to--nearIR imaging data analyzed in \citet{Calzetti2023} (Figure~\ref{fig:Field}), but adopt a more general selection process than those authors to recover a larger sample of young star clusters, with a greater diversity of ages, masses and extinctions. We select our sources based on the presence of ionized gas emission identified from the infrared hydrogen recombination line Pa$\beta$ ($\lambda$1.282~$\mu$m); this ensures that our sources are typically young, $\lesssim$6--7~Myr, since beyond 7~Myr the ionized gas emission is too faint to be detected  \citep{Leitherer+1999}. However, we do not impose explicit constraints on mass or extinction, thus mitigating selection biases on these parameters. 

The paper is structured as follows: section \ref{sec:Image Data} contains a description of the imaging data, section \ref{sec:Source Catalog} describes the construction of the source catalog, and section \ref{sec:JWST} details the cross--validation of the sources using a JWST image of NGC\,4449. Section~\ref{sec:Photometry} describes the multi--wavelength flux  measurements using aperture photometry for the sources. The stellar population and dust attenuation models are described in section~\ref{sec:Models and Fitting}. Results and discussion are in section~\ref{sec:discussion}, while the summary and conclusions are in section~\ref{sec:summary}.
 
\section{Imaging Data} \label{sec:Image Data}

The data used in this survey is comprised of %\emph{Hubble Space Telescope} (HST) 
HST imaging covering from nearUV (NUV) to nearIR (NIR) in 10 bands. With the WFC3/UVIS and ACS instruments, we sample the NUV to optical range with broad-band filters (F275W, F336W, F435W, F555W and F814W), a H${\alpha}$+[NII] narrow-band filter (F657N), and a medium-band filter sampling the line-free continuum at V (F550M) (Table \ref{tab:Targets}).  Further, with WFC3/IR we sample the NIR with two broad-band (F110W and F160W) and one narrow-band (F128N) filter to cover the Pa${\beta}$ hydrogen recombination line emission. In summary, the 10 filters spanning the NUV--NIR include two hydrogen recombination lines (Table \ref{tab:Targets}). For each IR filter, the standard calibration pipeline CALWFC3 v. 3.5.2 was used to process individual frames into the final images, after correction for bias, dark, and flat fields \citep{Messa2021}. The NUV and optical images were retrieved from the MAST\footnote{Mikulski Archive for Space Telescopes, https://archive.stsci.edu/missions-and-data/hst} archive already processed through their respective instrumental pipelines. All images were aligned and mosaiced to the Gaia DR2 reference frame and then resampled to 0.04''/pix (NUV/optical) or 0.08''/pix (NIR), using DrizzlePac. The NIR images are resampled to twice the pixel size of the NUV/optical images because the Point Spread Function (PSF) of the WFC3/IR images is twice that of the optical and UV images. At the distance of NGC\,4449, the pixel scale of  0.04'' corresponds to 0.81~pc. Using the header keyword PHOTFLAM, images were then converted from instrumental units (e$^{-}$/s) to physical flux units (erg/s/cm$^2$/\AA). Table \ref{tab:Filters} contains each filter's respective pivot wavelength and FWHM. Since the WFC3/IR images have the narrowest field of view of the sample, the source search area of NGC\,4449 is limited by their coverage (Figure~\ref{fig:Field}). This area covers the central 2.8$\times$2.4 kpc$^2$ of NGC\,4449 and captures 67\% of the total SFR of the galaxy.

We produce emission line images by subtracting stellar continuum from the narrow-band images, F658N and F128N. For the F658N image (H$\alpha$+[NII]), we use the F550M and F814W images to perform an interpolation of the continuum flux. For the F128N (Pa$\beta$) image, we use the F110W and F160W images for the interpolation and the resulting continuum maps are then scaled using isolated foreground stars within the images which do not have a contribution from any nebular emission. 
The presence of Pa$\beta$ in the F110W filter requires the subtraction to be performed iteratively to adequately remove the continuum from the final line image. After producing the first F128N line emission map, we subtract this map from the original F110W image to create a line-free 1.1um continuum image. We then repeat the interpolation, scaling, and subtraction process until the final image differs from the previous version by less than 1\%; this is achieved with two iterations.  The continuum subtracted images are then multiplied by their filter widths, given by their full width at half maximums, and corrected for the filter's transmission at the galaxy's redshift to obtain line fluxes. The line emission in the F658N image is corrected for the [NII] contribution  using a [NII]/H$\alpha$ of 0.11 for NGC\,4449 \citep{Berg+2012}.

%We also create a second Pa$\beta$ image at twice the pixel scale (0.08''/pix) since the Point Spread Function (PSF) of the WFC/IR images is twice that of the optical and UV images. The Pa$\beta$ image with twice the pixel scale is used to identify false Pa$\beta$ sources that may be caused by slight misalignment between the three NIR images and to perform aperture photometry.

\begin{deluxetable}{cccc}
\centering 
\tablecolumns{4}
\tablewidth{0pt}
\tablecaption{Data Sources}
\tablehead{Instrument$^{1}$ & Filters$^{2}$ & Exposure Times$^{3}$ & Proposal ID$^{4}$}
\label{tab:Targets}
\startdata
WFC3/UVIS & F275W, F336W & 2481, 2360 & 13364\\ 
ACS/WFC & F435W, F550M, F555W, F658N, F814W & 7140, 1200, 4920, 2260, 4660 & 10522, 10585\\ 
WFC3/IR & F110W, F128N, F160W & 1000, 2600, 1700 & 15330 \\
\enddata

\tablenotetext{1}{WFC3/UVIS=Wide Field Camera 3 UV–Visual channel. ACS/WFC= Advanced Camera for Surveys Wide Field Channel. WFC3/IR=Wide Field Camera 3 Infrared channel.}
\tablenotetext{2}{HST Filter names.}
\tablenotetext{3}{Exposure time for filter in seconds.}
\tablenotetext{4}{ID number of the GO program that obtained the images: GO–13364 (LEGUS, Legacy ExtraGalactic UV Survey), PI: Calzetti; GO–10522, PI: Calzetti; GO–10585, PI: Aloisi; GO–15330, PI: Calzetti.}
\end{deluxetable}

\begin{deluxetable}{cccc}
\centering 
\tablecolumns{4}
\tablewidth{0pt}
\tablecaption{Filter Table}
\tablehead{HST Filter & Pivot $\lambda$ & FWHM  & Standard Filter \\ {} & $\mu$m  & $\mu$m & {}}
\label{tab:Filters}
\startdata
F275W & 0.2710 & 0.04053 & NUV\\ 
F336W & 0.3355 & 0.05116 & U\\
F435W & 0.4329 & 0.06911 & B\\
F550M & 0.5581 & 0.03845 & Medium V\\
F555W & 0.5360 & 0.08478 & V\\
F658N & 0.6584 & 0.00875 & H$\alpha$+[NII]\\
F814W & 0.8048 & 0.15416 & I\\
F110W & 1.1534 & 0.44300 & J\\
F128N & 1.2832 & 0.01590 & Pa$\beta$\\
F160W & 1.5369 & 0.26830 & H\\
\enddata
\end{deluxetable}

\section{Source Catalog} \label{sec:Source Catalog}

Within the area of NGC\,4449 determined by the field of view of the NIR pointing (Figure~\ref{fig:Field}), we identify sources that have compact Pa${\beta}$ recombination line emission, therefore, are potentially young star clusters. This approach is similar to the one employed by \citet{Messa2021}. Those authors use a second method for specifically isolating dusty sources by searching for high--extinction compact sources in extinction maps. Here we only concentrate on the compact Pa$\beta$ method, because we are interested in isolating dusty sources that are also young. There is considerable overlap between the catalogs selected with the two methods: 65\% of the sources from the extinction map method are in common with the sources from the compact Pa$\beta$ method; the latter method is also very efficient, since it retrieves 84\% of the sources in common between the two catalogs \citep{Messa2021}.

Star cluster candidates are extracted from both the ACS F814W image and the continuum-subtracted Pa${\beta}$ line emission image at their respective pixel scale (0.04''/pix and 0.08''/pix, respectively). We use SExtractor \citep{Bertin1996} to perform the source extraction. Input parameters are optimized to identify sources with a $\geq$5$\sigma$ detection in 5 and 3 contiguous pixels in the F814W and F128N, respectively. This identifies 224,342 and 1240 sources in the F814W and Pa${\beta}$ images, respectively. We then match the two catalogs within a 2 pixel (0.08" or 1 pixels in the IR channels) tolerance, retaining 1119 common sources. Visual checks were performed to ensure that the low tolerance was not discarding potential matches. This initial catalog is then inspected by eye, using the full suite of images.
%sampled at 0.04''/pix and the IR images sampled at 0.08''/pix. 
Apart from image artifacts and edge detections, the main source of contamination in this catalog are poorly--subtracted sources which appear as detections in our Pa${\beta}$ line map with Signal--to--Noise ratio S/N$\geq 5$ relative to the locally--measured background. In regions where the stellar continuum is over(under)--subtracted, the sources that are not perfectly subtracted will be preferentially removed (retained) in our catalog. 
These sources are straightforward to visually identify and remove from the catalog as they include negative pixel values and live in regions where the overall background is over-subtracted. Since these false detections arise due to poor continuum subtraction rather than genuine Pa$\beta$ emission, omitting them does not bias our catalog against additional young clusters, nor does the issue of over-subtraction mask any real sources. We also remove sources whose peak Pa$\beta$ emission does not appear coincident with the peak stellar continuum emission, since these may be the result of several sources in superposition in a compact association or may be slightly evolved sources that have ejected their surrounding gas. After this visual inspection we retain 464 sources. 

\section{Confirmation of Sources with JWST Data}\label{sec:JWST}

We cross reference the locations of the remaining 464 sources to a stellar--continuum subtracted Pa$\alpha$ ($\lambda$ 1.8756~$\mu$m) line emission map acquired with JWST. The JWST Pa$\alpha$ image has been obtained using the NIRCam F187N filter, continuum--subtracted using adjacent broad--band filters (private communication, A. Adamo 2023). This image has double the angular resolution and is deeper than our HST/WFC3/IR images, thus providing a sanity check for our nebular emission detections. The Pa$\alpha$ hydrogen recombination line serves as an additional marker for young star clusters while being less susceptible, by virtue of being at longer wavelength, to the effects of dust attenuation than  Pa$\beta$. We perform a visual inspection of the Pa$\alpha$ image at the location of our sources, discarding all sources that may be poorly subtracted stars. In addition, we impose a detection limit cut--off and only deem acceptable those sources that are detected in Pa$\alpha$ with S/N$\ge$3. This last criterion removes 167 faint emission line sources, which may be either old clusters or young low--mass ones. We do not expect that discarding low--mass young clusters will affect our conclusions, which are based on sources with mass$\ge$3,000~M$_{\odot}$ (see section \ref{sec:discussion}). We note that we do not recover any of the sources identified by \citet{Calzetti2023}; none of the 34 sources selected by those authors show evidence for emission in Pa$\alpha$, although they are sufficiently compact to be indistinguishable from stellar sources and therefore part of the stellar population in this galaxy. After applying both of our selection criteria, we are left with 99 sources (21\% of the original sample) that display Pa$\alpha$ emission, with $>$90\% of them detected with S/N$>$7 in the line. This is the sample our analysis centers on.

\section{Photometry} \label{sec:Photometry}

Aperture photometry is performed on each of the 99 cluster candidates. We adopt a fixed 3/1.5 (NUV+optical/NIR) pixel radius (0$''$.12=2.44 pc) to ensure that photometry is not affected by nearby contaminants. A local sky annulus in the range 5-7/2.5-3.5 (NUV+optical/NIR) pixels is used for the  background subtraction of each source. Prior to computing the photometry, each source is centered on the F814W using a centroiding algorithm and fixed in all filters to this point. The F814W is used as reference since it is our reddest optical band, with a point spread function (PSF) roughly half the size of our NIR bands. This makes the F814W ideal as a reference since it can capture the flux of highly extincted sources compared to shorter bands while still having sufficient angular resolution to differentiate between the source and nearby contaminants.

We account for flux outside of the fixed aperture by applying aperture corrections which were determined using the methods described in \citet{Messa2021}. Briefly, to estimate the corrections, we convolve isolated stellar sources in each band with a Moffat profile, considered to be the most accurate light profile function for describing clusters \citep{Bastian+2013}. We use a R$_{\textrm{eff}}$ value of 2.5 pc, which is the peak of the cluster radius distribution \citep{Ryon+2015,Ryon+2017}, to obtain an average correction in each band for all clusters. Although the choice of R$_{\textrm{eff}}$ affects the normalization of the SED, it does not affect its overall shape. Thus, when fitting the SEDs to models, the only physical property that is affected is the resulting mass of the cluster. In addition, \citet{Messa2021} found that the resulting change in mass distribution between R$_{\textrm{eff}}$ values of 1.3 pc (the average radius of clusters in their sample) and 2.5 pc is minimal. We apply the following corrections (all expressed in AB mag):  -1.54±0.09 for F160W, -1.52±0.09 for F128N, -1.39±0.09 for F110W, -1.19±0.08 for F814W, -1.16±0.08 for F658N, -1.13±0.08 for F555W, -1.08±0.08 for F550M, -1.13±0.08 for F435W, -1.03±0.08 for F336W, -1.04±0.08 for F275W. Finally, we correct the photometric measurements for foreground galactic extinction with the MW foreground extinction, E(B–V)=0.017. Total uncertainty is derived from a combination of photon noise, uncertainty in aperture corrections, and the standard deviation of the background. We check for relative (band--to--band) photometric accuracy between broad and narrow--band filters by measuring several stellar sources in the field of NGC\,4449. The stellar sources do not show, or in a few cases only have non--significant ($<$1~$\sigma$), emission lines, either in H$\alpha$ or Pa$\beta$, indicating that the narrow--band filters are accurately calibrated within a few percent, and we should not expect to detect `spurious' emission lines due to mis--calibrations. Photometry of the 99 sources is listed in Table~\ref{tab:Source Table} and their location is shown in Figures~\ref{fig:Field} and \ref{fig:4 NGC4449 plots}. 

\begin{figure}
    
    \centering
    \includegraphics[width = 0.8\linewidth]{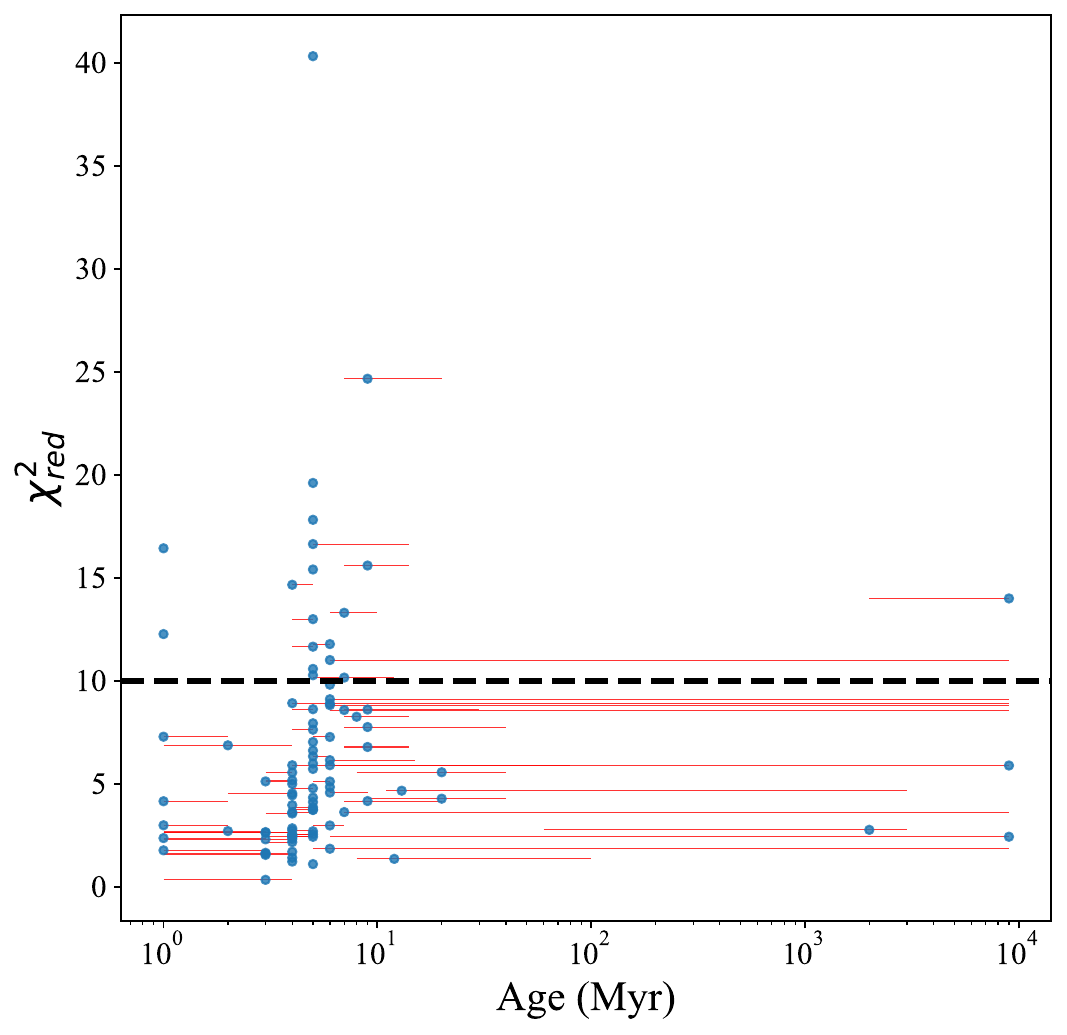}
    \caption{Minimum ${\chi}^2_{red}$ versus age of 99 fitted sources. The black dashed line denotes a $\chi^2_{red}<10$ cutoff below which there are 80 sources. The 1$\sigma$ uncertainties in age are calculated from a 1$\sigma$ interval from the minimum ${\chi}^2_{red}$.}
    \label{fig:ChiSq distribution}
\end{figure}

\begin{figure}
    
    \centering
    \includegraphics[width = 1.0\linewidth]{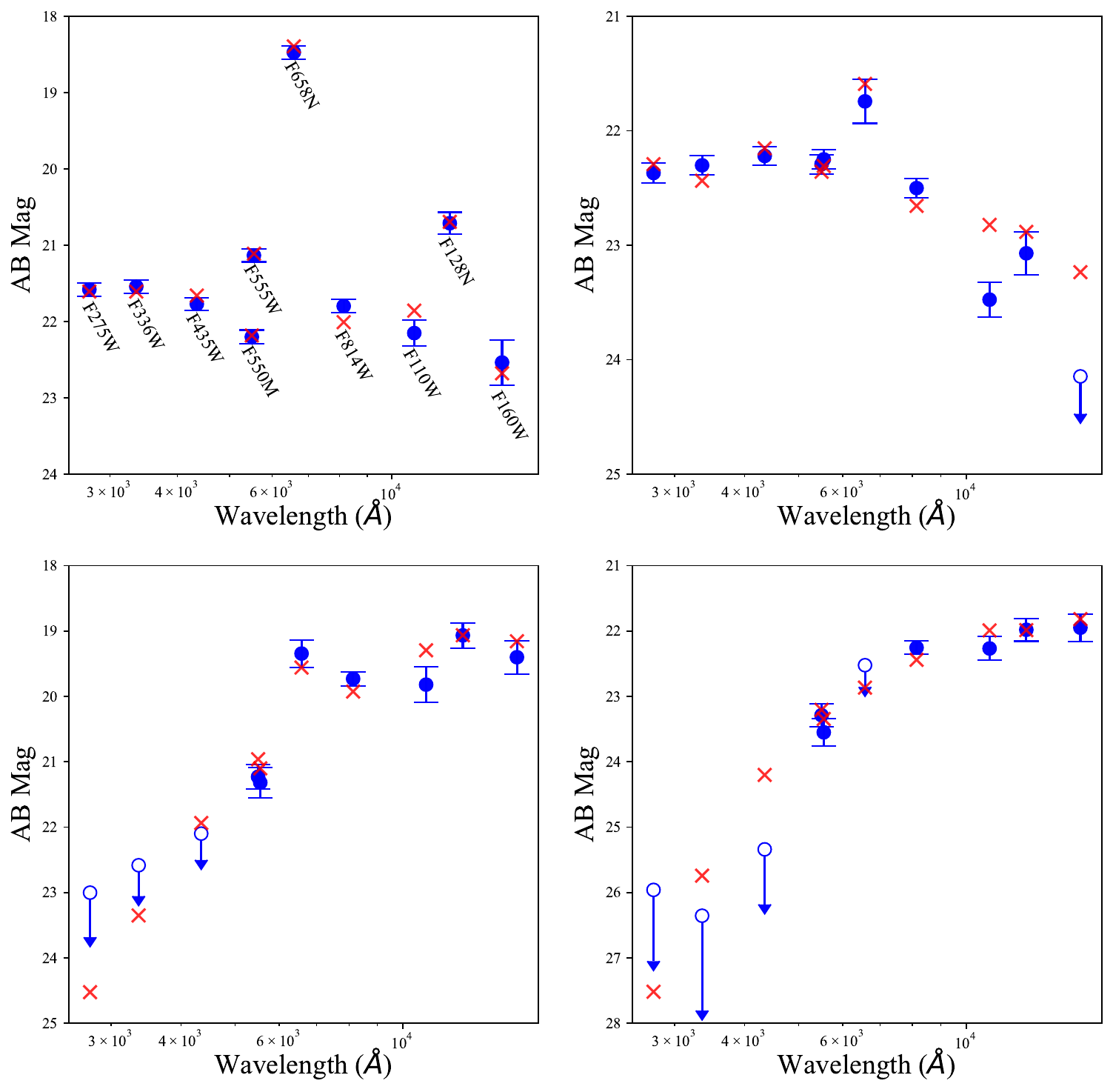}
    \caption{4 SEDs of sources with $\chi^2_{red}<10$ exhibiting a range of ages and extinctions. The red X marks the best fit value. Open circular markers are upper limits and solid circular markers are detections. (Upper Left) 1 Myr/E(B-V)=0.12mag. (Upper Right) 6 Myr/E(B-V)=0.0mag. (Lower Left) 6 Myr/E(B-V)=1.1mag. (Lower Right) 9000 Myr/E(B-V)=0.04mag.}
    \label{fig:SED plot}
\end{figure}

\section{Models and Fitting} \label{sec:Models and Fitting}
We fit our photometry to synthetic models to derive physical properties of the cluster candidates including age, mass, and extinction. We utilize Yggdrasil SED models \citep{Zackrisson+2011} in combination with dust attenuation/extinction models to produce synthetic photometry similar to the procedure followed in \citet{Calzetti+2015b}, \citet{Adamo+2017}, \citet{Messa2021}, and \citet{Calzetti2023}. 

Yggdrasil generates single stellar population (SSP) models by utilizing the Starburst99 \citep{Leitherer+1999} spectral synthesis models and then combining these models with Cloudy \citep{Ferland+2013}, to generate the nebular emission lines \citep{Zackrisson+2011}. The Starburst99 models already include both stellar and nebular continuum. 

The SSP models we use are generated assuming instantaneous star formation, a Kroupa Initial Mass Function (IMF) \citep{Kroupa2001} in the range 0.1-120 M$_\odot$, and metallicity of Z=0.008, consistent with the metallicity of NGC\,4449, which is about 40\% of the solar value \citep{Berg+2012}. Ages are sampled in 1 Myr increments between 1 Myr and 10 Myr, with increasingly sparser age sampling out to 10~Gyr.  We discuss later the impact of stochastic sampling of the stellar IMF on our source parameters determinations \citep{Cervino2002}.

The SSP SEDs are attenuated with both a starburst attenuation curve \citep{Calzetti+2000} and an LMC extinction curve \citep{Fitzpatrick1999}. We use a foreground dust geometry \citep{Calzetti2001} of the form:

\begin{equation}
F(\lambda)_{final} = F(\lambda)_{model} 10^{[-0.4 E(B-V) \kappa(\lambda)]},
\end{equation}
where E(B-V) is the color excess and $\kappa(\lambda)$ is the attenuation/extinction curve. The models range in E(B-V) from 0--5 mag with increments of 0.02~mag. The dust-attenuated SED models are then convolved with the HST filters' transmission curves to produce synthetic photometry. 

We then fit the SEDs derived from aperture photometry to these synthetic photometry models using ${\chi}^2$-minimization, taking into account the uncertainties in each band as well as upper limits. A source must possess at least 3 bands with $\sigma$$<$0.3 for the source to be fit. To account for highly extincted sources with weak, highly uncertain detections at shorter wavelengths, we utilize the uncertainty in the F555W as a reference point. If $\sigma_{F555W}$ exceeds 0.3 mag, we treat the F555W and all shorter wavelength bands as upper limits during the fitting process. This approach resolves the issue of poor fits caused by a lack of detection at shorter bands. With this criteria we were able to successfully fit all of the 99 sources. 

Fitting returns a distribution of reduced-$\chi^{2}$ ($\chi^2_{red}$) values for each model and its corresponding physical properties. We take the model with the minimum $\chi^2_{red}$ to be the best fit. With this, we find that the models with the LMC extinction curve produce better fits than the models with the starburst attenuation curve, in agreement with the findings of \citet{Calzetti2023}. We thus adopt the models with the LMC extinction as our reference models. We determine an uncertainty interval for the best-fit parameters of each source to be all fits within 1$\sigma$ of the minimum $\chi^2_{red}$ value, a corresponding $\chi^2_{red}$ interval of 3.53 based on our 3 degrees of freedom. The list of cluster properties, ages, masses and extinctions, with 1~$\sigma$ uncertainties and the value of the minimum $\chi^2_{red}$ from the fits is given in Table~\ref{tab:Source Properties Table} for all 99 sources.

\begin{figure}
    
    \centering
    \includegraphics[width = 1.0\linewidth]{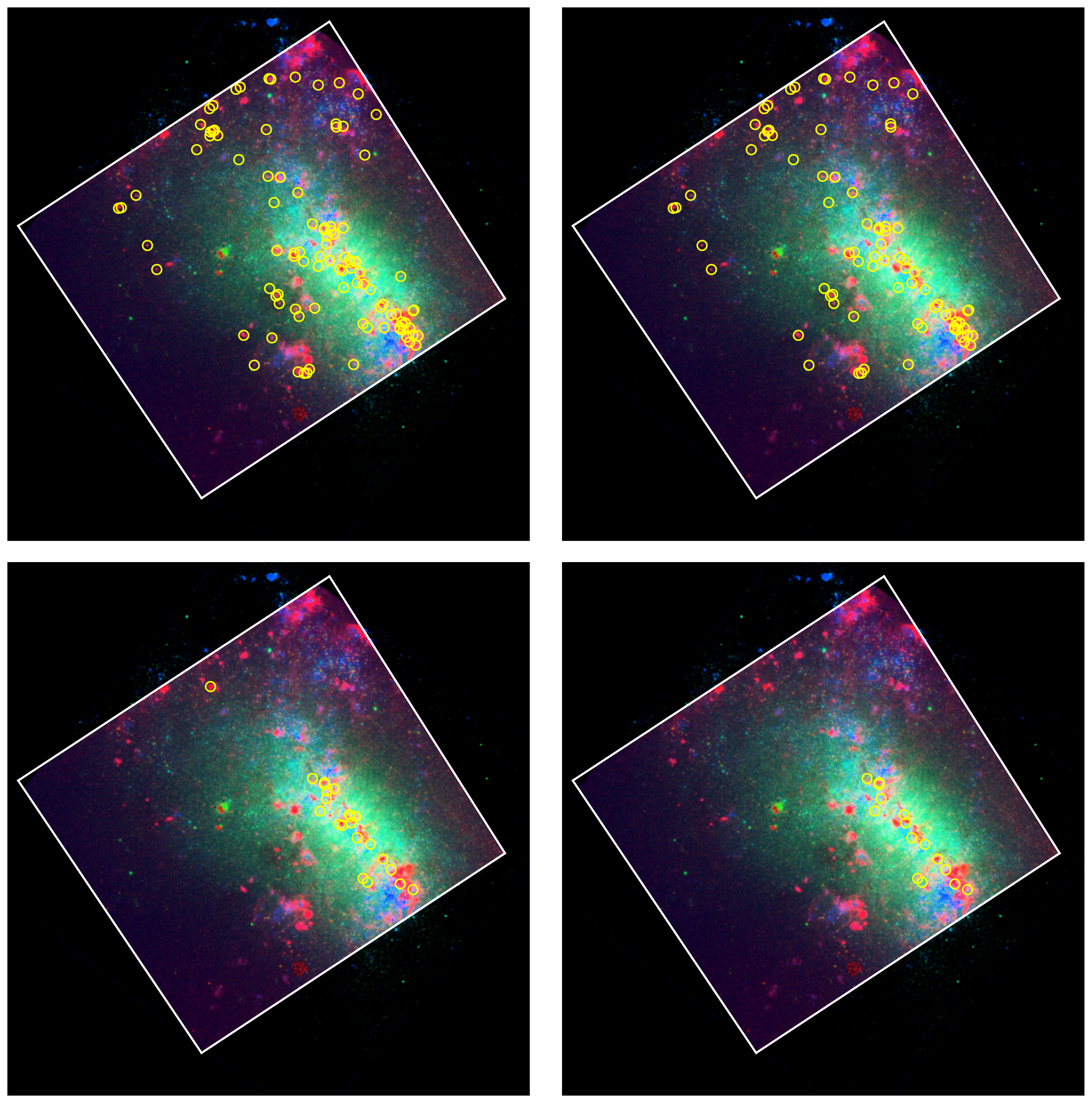}
    \caption{Three color composites of NGC\,4449 showing each stage of the sample selection, with the sources marked with yellow circles. (Upper Left) Full set of successfully fitted sources (99 sources). (Upper Right) Successfully fitted sources with $\chi^2_{red}<10$ (80 sources). (Lower Left) Successfully fitted sources with M$>$3000~M$_{\odot}$ (22 sources). (Lower Right) Successfully fitted sources with both $\chi^2_{red}<10$ and M$>$3000~M$_{\odot}$ (15 sources).}
    \label{fig:4 NGC4449 plots}
\end{figure}

Figure \ref{fig:ChiSq distribution} illustrates the resulting distribution of best-fit ages and minimum $\chi^2_{red}$ of our sample. We define a minimum $\chi^2_{red}$ cutoff of 10 to ensure a well-fit sample, following the approach of \citet{Adamo+2017}. Of the 99 sources with sufficiently accurate photometry to be fit with our method (see above), 80 sources possess $\chi^2_{red}<10$. Figure \ref{fig:SED plot} illustrates four SEDs and their best-fits with $\chi^2_{red}<10$. These four sources span the whole parameter space of age and color excess in our sample. The location of the 80 sources with $\chi^2_{red}<10$ is shown in Figure~\ref{fig:4 NGC4449 plots} (upper--right panel).

\section{Results and Discussion}\label{sec:discussion}

We analyze the correlation between physical properties obtained from the SED fitting, presenting samples selected with different cuts, and discuss their implications. 

\begin{figure}
    
    \centering
    \includegraphics[width = 1.0\linewidth]{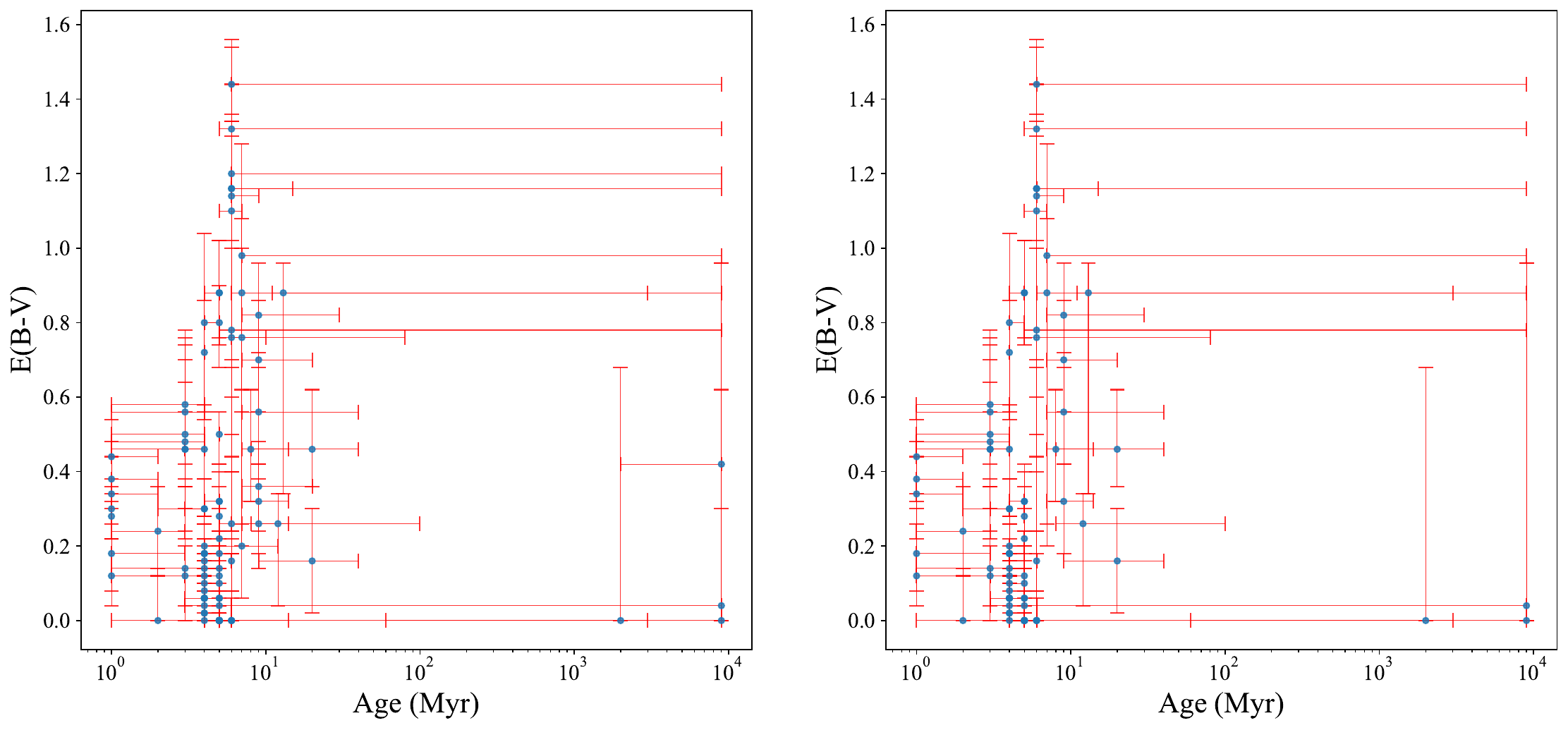}
    \caption{Best fit age versus E(B-V). (Left) All 99 fitted sources. (Right) The 80 sources with $\chi^2_{red}<10$.}
    \label{fig:Age Color}
\end{figure}

Figure \ref{fig:Age Color} illustrates the age and color-excess relation of the sources in our sample, using both the entire sample of 99 sources (left) and the sub--sample of 80 sources with $\chi^2_{red}<10$ (right). We observe a plume of sources ranging in E(B-V) from 0~mag to $\sim$1.4~mag centered at 5-7 Myr. We also observe a limited number (16 total, 14 with $\chi^2_{red}<$10) of sources with ages $\le$3 Myr, none of which possesses E(B-V) exceeding 0.56 mag. The overall shape of both the full sample and the one with  $\chi^2_{red}<10$ are qualitatively similar. 

\begin{figure}
    
    \centering
    \includegraphics[width = 1.0\linewidth]{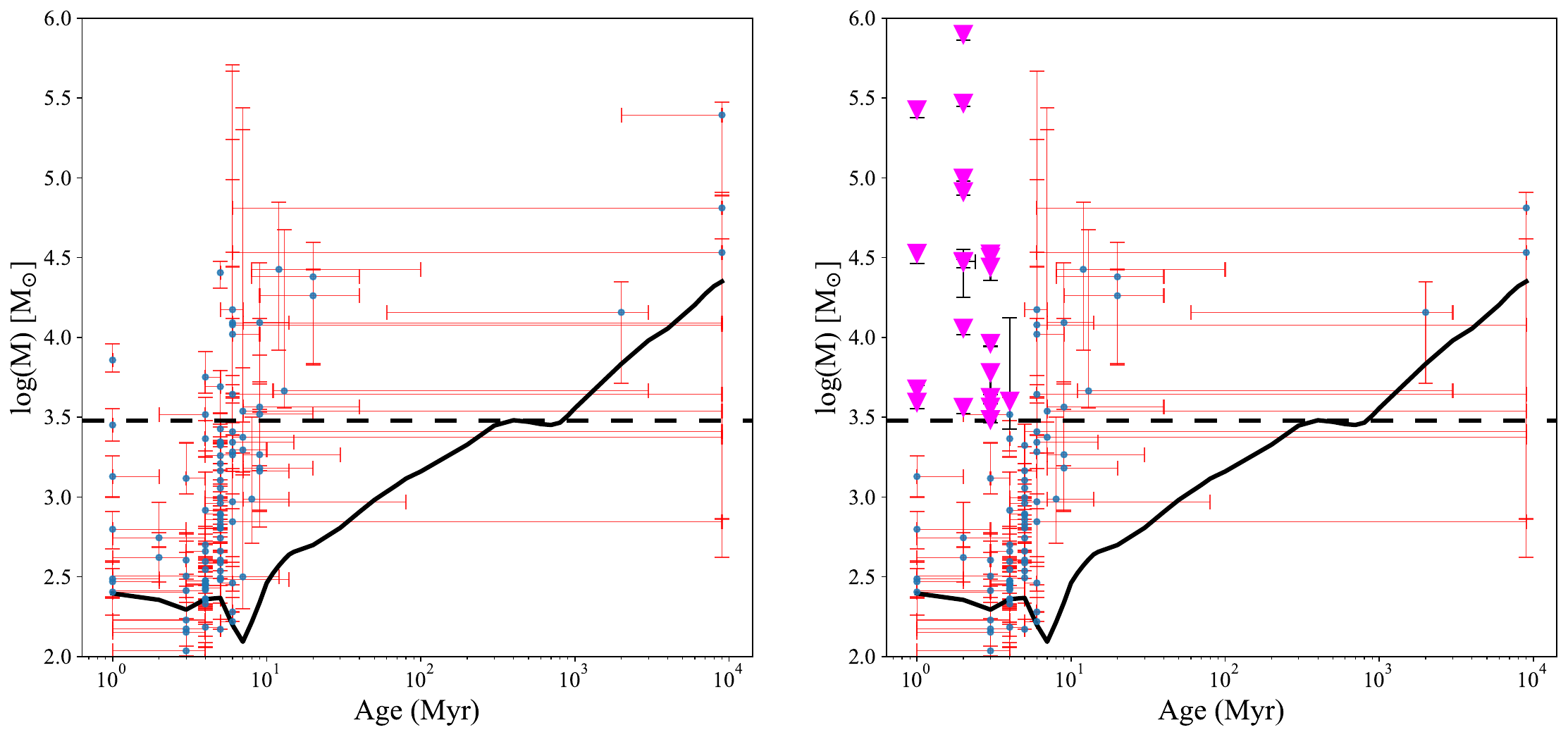}
    \caption{Mass limit (black solid line) as a function of age, as determined from the I-band detection limit of our images. 
    %given by the lowest mass cluster in our sample, $\sim 10^{2}$ M$_{\odot}$. 
    We set a 3000 M$_{\odot}$ lower limit (black dashed line) to avoid impact on our physical parameters from stochastic sampling of the IMF (see Section \ref{sec:Models and Fitting}). (Left) Mass versus age for all 99 fitted sources. (Right) The same plot as to the left, only for the sources with $\chi^2_{red}<$10. In the right panel, we also show the location of the LEGUS star clusters with $\chi^2_{red}<$10,  mass$\ge$3,000 M$_{\odot}$ and coincident with ionized gas emission  (magenta triangles). See \citet{Adamo+2017} for a discussion of the LEGUS star clusters.}
    \label{fig:Age Mass}
\end{figure}

The age--mass plots (Figure \ref{fig:Age Mass}) show that the 5-7 Myr plume has best-fit masses in the range  $\sim$10$^2$-10$^{4.2}$ M$_{\odot}$ with higher masses, up to 10$^{5.5}$ M$_{\odot}$, for the few  $>$20~Myr sources. The low mass end is consistent with the mass limit we can reach given the depth of our images. We derive the mass limit as a function of age from the I-band detection limit, and its track is shown in Figure \ref{fig:Age Mass}.
%of our lowest mass source ($\sim$100~M$_{\odot}$). 

\begin{figure}
    
    \centering
    \includegraphics[width = 1.0\linewidth]{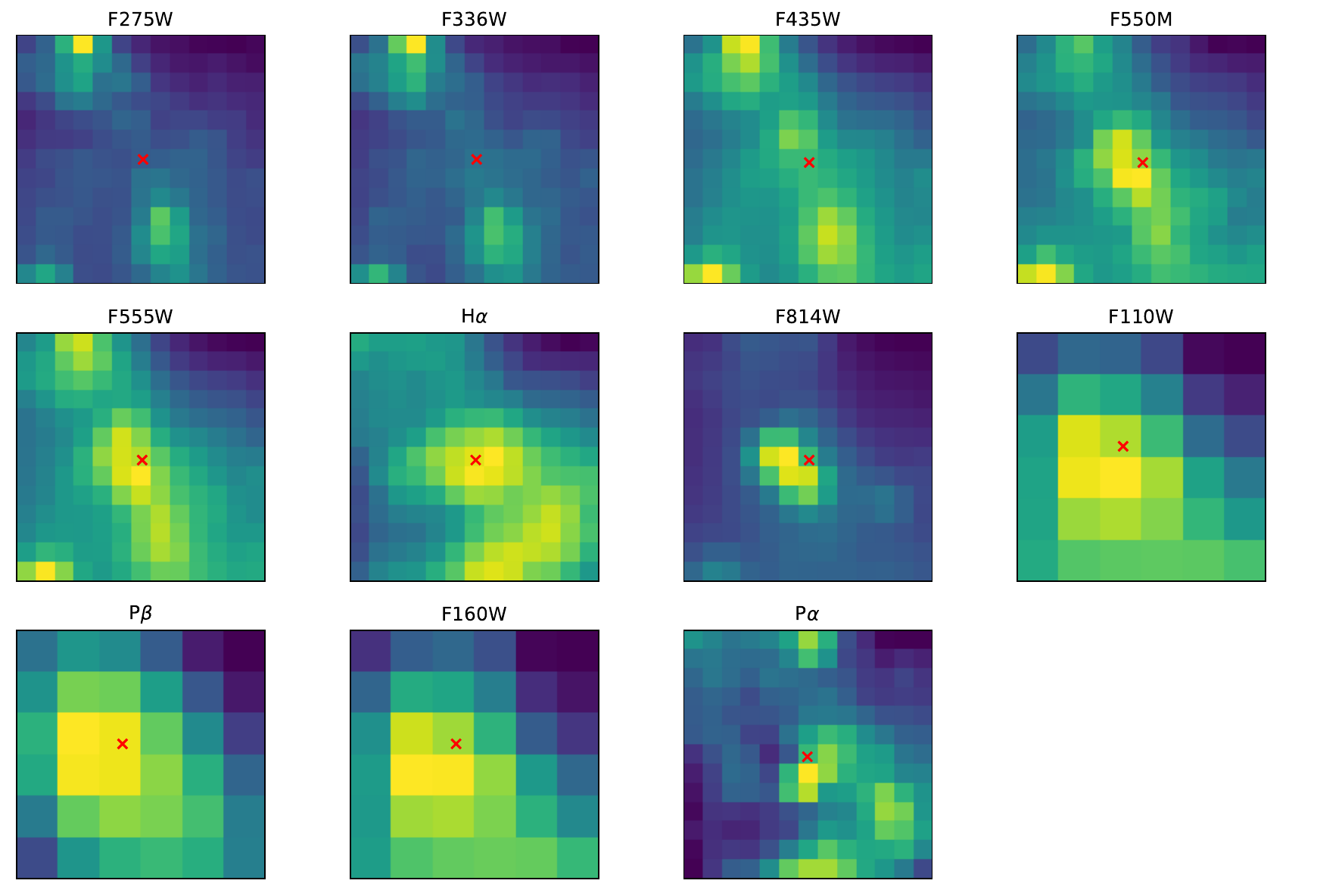}
    \caption{Cutouts corresponding to the location of source ID 82 with a best fit E(B-V) of 1.1 mag and an age of 6 Myr whose SED is displayed in the lower left of Figure \ref{fig:SED plot}. H$\alpha$, P$\beta$, and P$\alpha$ are continuum-free emission line maps. Each box has dimensions of $0.5''\times0.5''$.}
    \label{fig:Source101}
\end{figure}

While young ($\lesssim$6~Myr), moderately dusty star clusters  tend to have strong Pa$\beta$ line emission, %(Figure~\ref{fig:Source101}), 
which is well modeled by our fitting routine (top panels of Figure \ref{fig:SED plot}), the fitting routine has difficulty discriminating between the SED of a dusty, $\sim$6--8~Myr old cluster (Figure~\ref{fig:Source101}) and the SED of a much older, but low dust, cluster, leading to some degeneracy in the fits; this is clearly illustrated in the two bottom panels of Figure \ref{fig:SED plot}, where a dusty 6~Myr SED (left) and an almost--dust--free 9000~Myr SED (right) are shown. Four sources in our sample (three with $\chi^2_{red}<10$) are older than 20~Myr and as old as $\sim$10~Gyr, according to their best fits. The requirement that sources be detected in Pa$\beta$ line emission with S/N$\geq$5 in SExtractor   is intended to break the degeneracy between "young and dusty" and "old and low--dust" sources. Further, a total of 10 sources with $\chi_{red}^2<$10 have large age uncertainties ($>$1~Gyr); these sources are all 6~Myr and older in age,  including the three sources with best fit ages $>$20~Myr, again demonstrating the limitations of the fitting routine when emission lines are weak. An additional consideration is that clusters around the 6--10~Myr age range can contain red supergiants whose brightness can over--shine their host cluster, further diluting evidence for a young stellar population \citep[see discussion in][]{Whitmore+2020, Calzetti2023}. Thus, the four sources with best fit age $>$20~Myr may be younger than the best fits indicate. For sake of simplicity, in what follows we neglect further consideration of the four sources with best fit ages older than 20 Myr, reducing our sample to 95 sources.

We test for biases in the SED fit results due to potential chance alignment of, e.g., a red supergiant star in front of an otherwise low--dust star cluster \citep[e.g.,][]{Whitmore+2020}. As red supergiants appear  around $\sim$6~Myr of age, this configuration would produce an unusually red SED which may be mistaken for a dusty one. We test this possibility by performing spot-checks on our sample: we measure the extinction values from the line emission ratio H$\alpha$/Pa$\alpha$ for three clusters with E(B$-$V)$>$1~mag (\# 13, 45, and 82), finding that the nebular emission yields E(B$-$V) values comparable to those obtained from the SED fits. We conclude that the results from the SED fits are robust in terms of the physical parameters they produce.

\begin{figure}
    
    \centering
    \includegraphics[width = 1.0\linewidth]{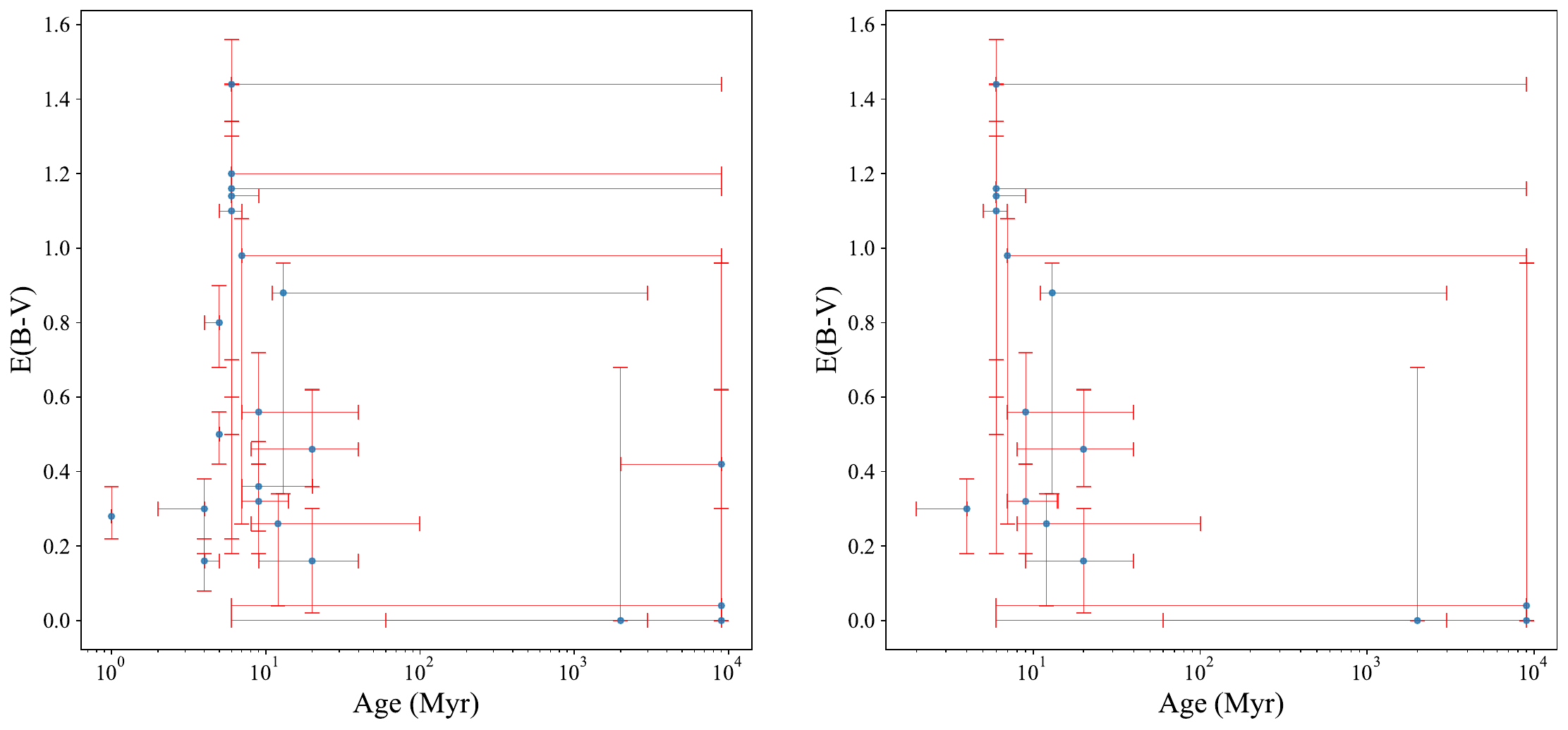}
    \caption{Age and color excess plots of all sources with masses $>$3000 M$_{\odot}$. (Left) 22 sources with any value of the $\chi^2_{red}$. (Right) 15 sources with $\chi^2_{red}<10$.}
    \label{fig:Age Color over 3000}
\end{figure}

To mitigate the effect of stochastic sampling of the IMF on the sources' derived physical parameters, we also define a 3000 M$_{\odot}$ mass cutoff which retains 22 sources with any $\chi^2_{red}$ and 15 sources with $\chi^2_{red}<10$ (Figure~\ref{fig:4 NGC4449 plots}). The age and color-excess of the resulting sample are presented in Figure \ref{fig:Age Color over 3000}. The 5-7 Myr plume is still visible, but the sample only contains three (one)  sources with age of 4~Myr or less. 

\subsection{Where are the clusters younger than 5~Myr?}
\begin{figure}
    
    \centering
    \includegraphics[width = 1.0\linewidth]{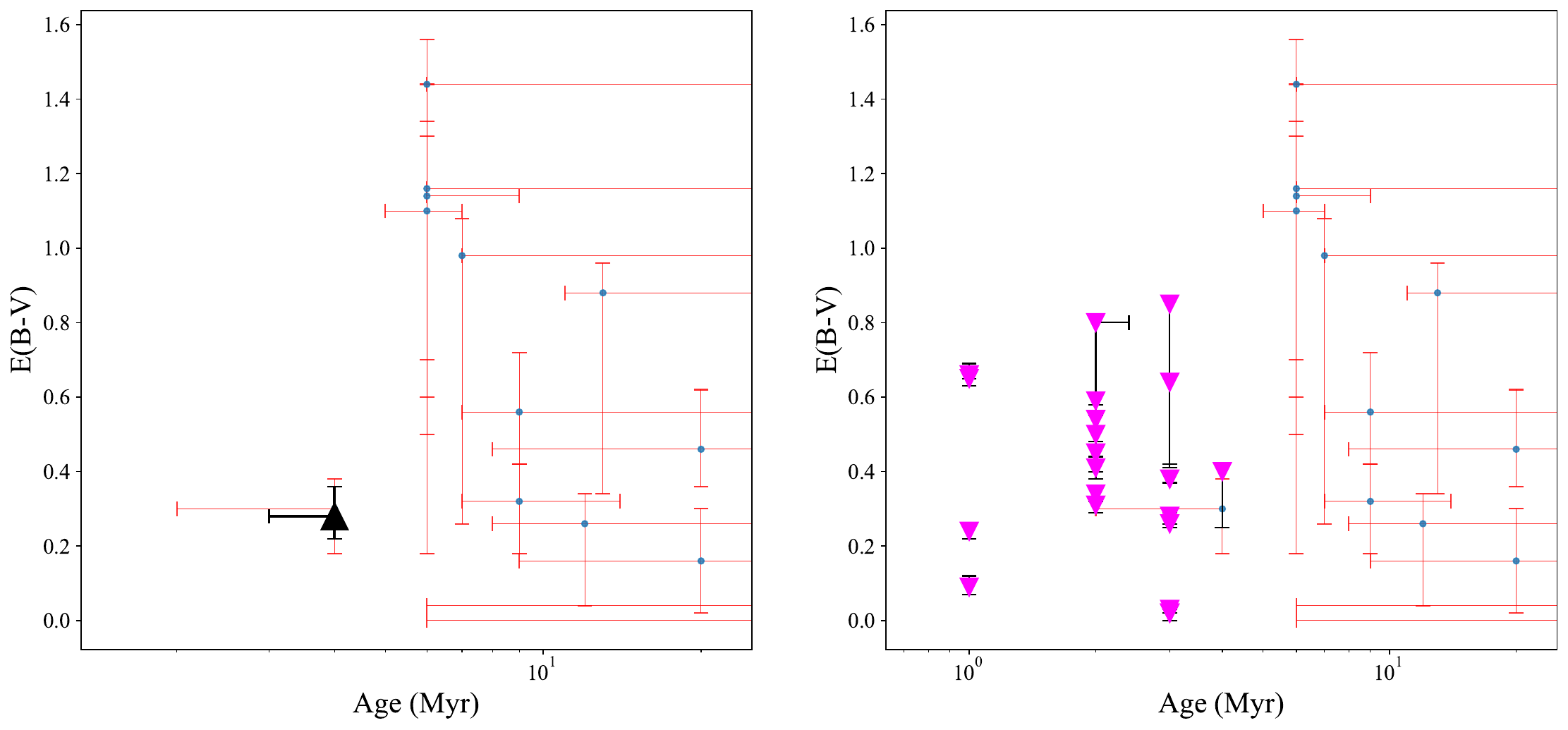}
    \caption{(Left) Best fit age versus color excess for the clusters with $\chi^2_{red}<10$ and mass $>$3000 M$_{\odot}$, and with the inclusion of the single visually identified source with comparable characteristics (black triangle). This is a relatively unextincted source with E(B-V)=0.28, suggesting that the approach of visual identification is insufficient at returning a sample of young, highly extincted clusters. (Right) The same clusters as in the left panel, with the addition of the LEGUS clusters with $\chi^2_{red}<10$, mass $>$3000 M$_{\odot}$ and coincident with ionized gas emission (magenta triangles with error bars). See \citet{Adamo+2017} for a discussion of the LEGUS star clusters.}
    \label{fig:Age Color 32 sources}
\end{figure}

Our source extraction parameters (detection in I-band and Pa$\beta$ line emission) were chosen to ensure an unbiased sample of young clusters. Despite this, we recovered relatively few very young ($\le$4~Myr) sources and just one with mass exceeding 3000 M$_{\odot}$ for the $\chi^2_{red}<$10 sample. In order to understand whether our selection method is biased against these very young and very dusty (E(B--V)$>1$~mag) clusters, we perform a by--eye search of line--emitting sources that are not already in our catalog. The requirement to have line emission ensures that the sources are young. In order to maximize the likelihood that the sources are also dusty, we perform the search around detected radio sources that are located within infrared--emitting regions. The radio sources are from the collection of \citet{Reines+2008}, who have identified compact radio--emitting regions in NGC\,4449 with an accuracy of $\sim$1$^{\prime\prime}$. Radio emission is less subject to dust extinction than optical and near--infrared wavelengths, so it can ensure an unbiased sample. The only shortcoming is that radio observations are limited to bright regions, i.e., massive star clusters. However, we search in the surroundings of these bright clusters for fainter (lower mass and/or more extincted) clusters. The infrared images we use to locate dusty regions are from the IRAC instrument on the \emph{Spitzer Space Telescope}, centered at 8~$\mu$m, a wavelength that targets dust emission; we use here the image presented in \citet{Calzetti+2018}. Our by--eye search yields 32 sources, only 18 of which are new, i.e., not already in our sample. Of these, 13 were successfully fit, and only a single source has a best-fit $\chi^2_{red}<10$ and mass $>$3,000~M$_{\odot}$, but it is relatively unextincted with an E(B-V)=0.28~mag. The location of this source on the E(B--V)--age plot is shown in Figure \ref{fig:Age Color 32 sources}. Of the remaining 12 fitted radio sources, 10 are low mass ($<$3,000~M$_{\odot}$) and 2 have poor $\chi^2_{red}$$>$15 and low E(B-V). These findings illustrate that our ability to recover both young and dusty sources in NGC\,4449 is not limited by our HST WFC3 UVIS/IR observations, whether through visual inspection or automated searches. 

We investigate whether our choice of models may drive some of these results. \citet{Whitmore+2020} compares the Yggdrasil models \citep{Zackrisson+2011} with  the Bruzual \& Charlot models \citep{Bruzual+2003}, finding  that the two sets of models generally yield comparable ages and masses. When discrepancies occur, they are in the direction of Yggdrasil assigning older ages ($>$40~Myr) to clusters that the Bruzual \& Charlot models yield young ages ($<$10~Myr) for. We believe this should not affect our sample, which is selected to have nebular line emission and, thus, young ages by default. 
To verify this, we fit a segment of our sources using the fitting code CIGALE \citep{Boquien+2019}, which includes the Bruzual \& Charlot models and nebular emission, both lines and continuum. We focus on the 11 sources with $\chi^2_{red}<10$, masses $>$3000 M$_{\odot}$, and ages between 5 and 20 Myr. The best fits we obtain from CIGALE indicate shifts towards marginally younger ages, down to 5 Myr, increased E(B-V) values up to 3 mag, and up to a $\sim$2.5 factor increase in mass. This test demonstrates that the fundamental picture remains consistent, marked by an absence of very young ($<$5 Myr), extincted sources.

%\textcolor{red}{To verify this, we fit a segment of our sources using the fitting code CIGALE \citep{Boquien+2019}, which includes the Bruzual \& Charlot models, focusing on all sources with $\chi^2_{red}<10$, masses $>$3000 M$_{\odot}$, and ages between 5 and 20 Myr (11 sources). The best fits we obtain from CIGALE indicate shifts towards marginally younger ages, down to 5 Myr, increased E(B-V) values up to 3 mag, and up to a $\sim$2.5 factor increase in mass. We therefore maintain that the fundamental picture remains consistent, marked by an absence of very young ($<$5 Myr), extincted sources.}

In addition, similar tests performed in \citet{Calzetti2023}, using the stellar population inference code Prospector \citep{Johnson+2021}, yielded no major changes or systematics in the ages/masses/extinctions of the sources relative to the results from Yggdrasil. In light of this discussion, we conclude that the choice of models/codes does  not have a major impact on our ages and extinction values. 

The considerations above leave us with two  possibilities: that compact and dusty star clusters with age $\lesssim$4~Myr are either absent in NGC\,4449 or have been somehow missed by our searches. To test either of these possibilities we investigate existing results on both the star formation history (SFH) of NGC\,4449 and the optically--selected young star clusters from the LEGUS project \citep{Calzetti+2015a}.

The SFH of NGC\,4449 has been derived by \citet{McQuinn+2010}, \citet{McQuinn+2012} and \citet{Sacchi+2018}, using resolved stellar populations detected in the ACS/WFC optical images from the Hubble Space Telescope, and by \citet{Cignoni+2018}, using both UV (WFC3) and optical (ACS/WFC) images. All authors recover similar shapes for the SFH, although there are differences in the details, especially in the youngest age bins. We concentrate on the results of \citet{Sacchi+2018}, as these authors include some of the most recent stellar evolution models in their analysis \citep[similarly to][]{Cignoni+2018} and divide the galaxy in sub--regions, helping a better match with our WFC3/IR footprint. Our footprint coincides with the `Center' region and part of the `Clump1' and `Clump2' regions identified by \citet{Sacchi+2018}. \citet{Sacchi+2018} finds that those central regions experienced a mild burst of star formation, that begun about 20~Myr ago, ended about 5--6~Myr ago and peaked around 14~Myr ago. Since the end of this burst, about 5--6~Myr ago, star formation in the central region of NGC\,4449 has been negligible, as inferred from resolved star populations. This result is consistent with the one obtained by \citet{Cignoni+2018} who recovered the SFH of the whole galaxy over the past $\sim$200~Myr. However, as the earliest stages of star formation produce clustered systems \citep[unresolved star clusters and associations, see][]{Lada+2003, Elmegreen2010, Elmegreen+2010} analyses that concentrate on resolved stars may miss such sources, and underestimate the SFH in the youngest age bins and specifically over the most recent few~Myr. 

To further test this explanation, we investigate the young star clusters identified in NGC\,4449 by the LEGUS project \citep{Calzetti+2015a} and discussed in \citet{Calzetti2023}. The LEGUS star cluster candidates were selected from UV,U,B,V,I images obtained with the WFC3/UVIS and ACS/WFC instruments on the HST. \citet{Whitmore+2020} added candidates to the NGC\,4449 cluster catalog based on inspection of the HST H$\alpha$ images. The requirement that a cluster needs to be detected in at least four bands in order to be attributed an age, mass and extinction from the SED fitting of its photometry constrains the selection to clusters that are detected in at least the UV (WFC3/F275W) or U (WFC3/F336W) \citep{Adamo+2017}. This sets also an upper limit to the maximum extinction of the clusters, determined by \citet{Whitmore+2020} to be E(B-V)$\lesssim$1~mag. The approach used by \citet{Adamo+2017} to the SED fitting is similar to the one adopted in this work, enabling a direct comparison. \citet{Calzetti2023} isolated 53 star cluster from the LEGUS catalog, with the following characteristics: mass$\ge$3,000~M$_{\odot}$, age$\le$10~Myr, and $\chi^2_{red}\le$10 from the SED fits. 

We inspected the location of these 53 star clusters on the narrow--band images H$\alpha$ and Pa$\beta$  to identify all LEGUS young clusters that are coincident with hydrogen recombination line emission, either compact or filamentary. We found only 21 such star clusters. The physical parameters (age, mass, extinction) of these 21 LEGUS cluster are plotted in Figures~\ref{fig:Age Mass} and \ref{fig:Age Color 32 sources}, where we include the small offset in mass due to the different choices in the distance of NGC\,4449 (3.9~Mpc for LEGUS versus 4.2~Mpc in this work). An interesting property of the LEGUS line--emitting clusters is that they are all $\le$4 Myr old and they extend to high masses, with median mass $\sim$3.5$\times$10$^4$~M$_{\odot}$. For comparison, the high--fidelity ($\chi^2_{red}\le$10) Pa$\beta$--emitting clusters we identify in our sample with mass $\ge$3,000~M$_{\odot}$ and age $\le$10~Myr (8~clusters) have ages $\gtrsim$5~Myr (with one at 4~Myr age) and median mass $\sim$7$\times$10$^3$~M$_{\odot}$, albeit with large uncertainties. Thus, our Pa$\beta$--emitting compact clusters are both systematically older and are a factor $\sim$5 less massive than the line--emitting clusters in the LEGUS sample. 

\begin{figure}
    
    \centering
    \includegraphics[width = 0.8\linewidth]{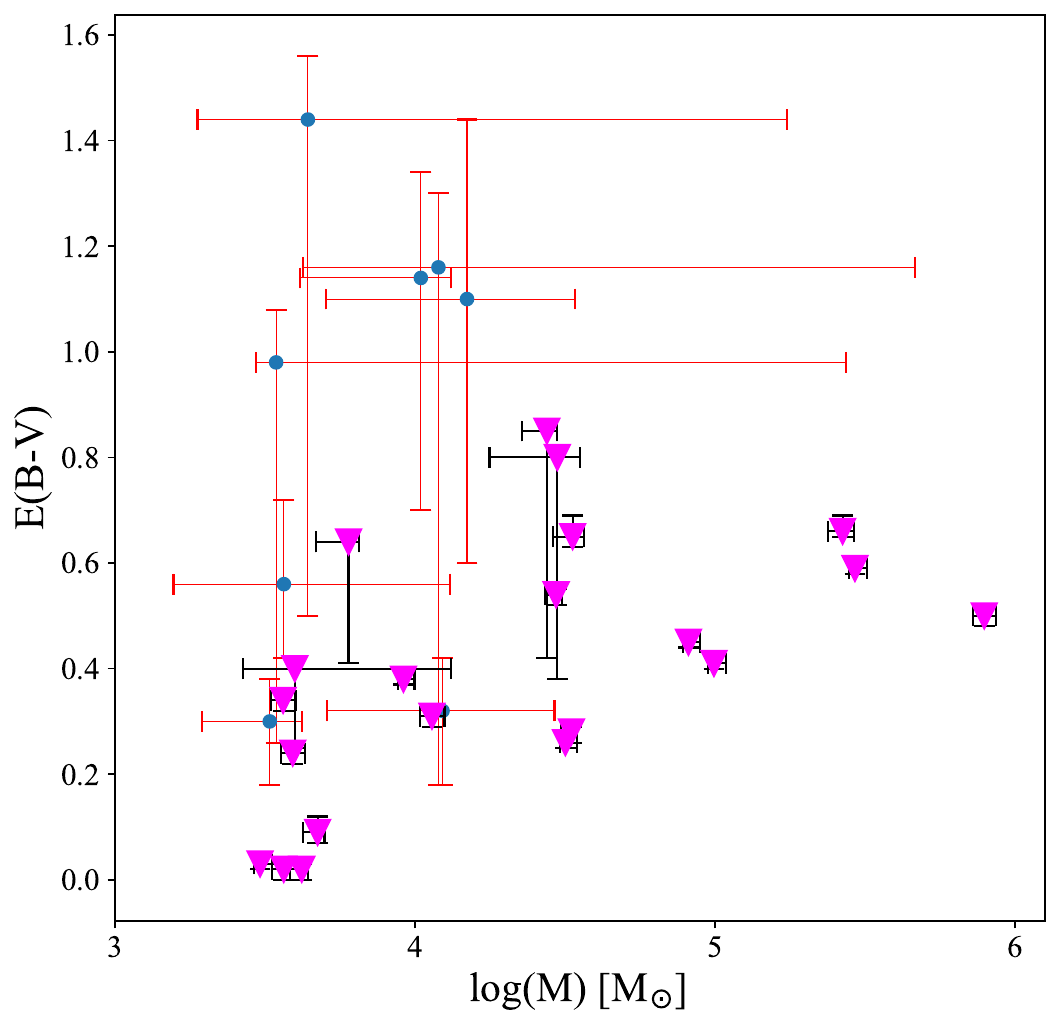}
    \caption{Color versus mass plot. Blue circles are sources with $\chi^2_{red}<10$, mass $>$3000 M$_{\odot}$, and age $\le$10 Myr (8 sources). Magenta triangles represent the sample of line--emitting LEGUS clusters.}
    \label{fig:Color mass}
\end{figure}

Figure~\ref{fig:Color mass} shows that the $\le$10~Myr old, Pa$\beta$--emitting sources are also on average  dustier than the line--emitting LEGUS sources, as expected from the selection criteria of each sample. While both sets of sources can have low extinction values, the LEGUS clusters have a maximum E(B$-$V)=0.85~mag, with a median value $\bar{E}(B-V)$=0.4~mag, while the Pa$\beta$--emitting sources in our sample have maximum E(B$-$V)=1.44~mag with median $\bar{E}(B-V)$=1.04~mag or a little over twice that of the LEGUS clusters. The age--dust degeneracy may slightly affect the mass assigned to a source if the age is incorrect. Much of this uncertainty is captured in the 1$\sigma$ uncertainties reported for our sources. For reference, in the age range 1--6.5~Myr, the mass inferred from the I--band varies by 0.33~dex and from the H--band by 0.19~dex, too small of a variation to affect the results presented in Figure~\ref{fig:Color mass}. 

In addition to differences in median age, mass, and extinction, the two samples of star clusters also differ in numbers. We find 21 line--emitting LEGUS clusters and 8 Pa$\beta$--emitting  clusters that are $\le$10~Myr old, $\ge$3,000~M$_{\odot}$ and with SED fits yielding $\chi^2_{red}<10$. However, when limiting the comparisons to the same mass range (3,000--15,000~M$_{\odot}$) for both samples, the numbers become comparable, since there are only 10 LEGUS clusters in this mass interval. Thus, there are no major discrepancies in the numbers recovered in the two samples, when the mass range is controlled for. 
One implication is that about half of the LEGUS clusters that are still retaining their gas are more massive than 15,000~M$_{\odot}$, but they require larger dust column densities ($\bar{E}(B-V)$=0.54~mag) than their lower mass counterparts ($\bar{E}(B-V)$=0.28~mag, Figure\ref{fig:Color mass}) to do so.

%Interestingly, these more massive clusters have higher extinction, about a factor of 2 higher or median $\bar{E}(B-V)$=0.54~mag, than the LEGUS clusters that are less massive than 15,000~M$_{\odot}$, which have median $\bar{E}(B-V)$=0.28~mag (Figure\ref{fig:Color mass}). Thus, among the clusters that emerge from their natal dust within $\sim$4~Myr, the massive ones are dustier than the lower mass ones. 

{\em Why have we missed the LEGUS young clusters in our catalog?} The line--emitting LEGUS star clusters are typically more extended and/or too faint in Pa$\beta$ than allowed by our selection criteria, %implemented in SExtractor, 
leading to these clusters being missed by our detection algorithm. In addition, many clusters are coincident with filamentary and/or diffuse line emission and are excluded by our matching criteria.

We now consider the sample of young, UV/optically--selected sources derived by \citet{Whitmore+2020}, who use an  approach to the SED fits that  differs from the default LEGUS one \citep{Adamo+2017}. In \citet{Whitmore+2020}, the five--band LEGUS photometry is fit with Bruzual \& Charlot models \citep{Bruzual+2003}, but without inclusion of nebular continuum or line emission; the authors add a prediction for the H$\alpha$ emission from the ionizing photons flux in the models, that they use to fit the photometry in the F658N filter. With these assumptions, \citet{Whitmore+2020} derive a slightly different set of young sources for NGC\,4449 from the one discussed above. We find 18 sources from the catalog of \citet{Whitmore+2020} that are good quality (flag Qual=1), younger than 10~Myr, more massive than 3,000~M$_{\odot}$,\footnote{We rescale the masses derived in \citet{Whitmore+2020} from their adopted distance of 4.31~Mpc to the 4.2~Mpc adopted here for NGC\,4449.} and are within the WFC3/IR footprint. All 18 sources are already present in the LEGUS catalog, but with several differences in the derived physical properties. Seven of the $>$3,000~M$_{\odot}$ clusters from the catalog of \citet{Whitmore+2020} have masses $<$3,000~M$_{\odot}$ in the LEGUS catalog. This is a direct consequence of the approach implemented by \citet{Whitmore+2020}: neglecting the nebular continuum at young ages yields overestimated stellar masses \citep{Reines+2010}. Of the remaining 11 clusters, seven have  derived ages $\sim$5~Myr in \citet{Whitmore+2020} and 10$\le$age$\le$30~Myr in LEGUS. Visual inspection of the H$\alpha$ image reveals that only one of the seven is truly associated with H$\alpha$ emission; the remaining six clusters are contaminated by neighboring strongly emitting line sources and do not appear to be associated with in--situ H$\alpha$ emission. The bona--fide H$\alpha$--emitting source is $\sim$5.5~Myr old, with mass $\sim$2.5$\times$10$^5$~M$_{\odot}$ and E(B--V)$\sim$0.44~mag. The remaining 4 sources (of 18) have derived ages $\sim$4--5~Myr in the catalog of \citet{Whitmore+2020} and $\sim$2--3~Myr in LEGUS, suggesting, again, that differences in the handling of the nebular continuum and lines may be at the root of the observed differences. The overall picture remains, however, unchanged, as already discussed: the young, UV/optically--selected  sources have ages$\lesssim$5~Myr and low extinctions when associated with line emission. We retain below the direct comparison with the LEGUS results, as our approach to SED fitting is similar to the one adopted in that project.

The picture that emerges from this analysis is that, in the central region of  NGC\,4449, star formation has continued for the past few Myr, producing mainly star clusters. Massive clusters retain their gas (and dust) for the first $\sim$4~Myr and then disperse it. Lower mass clusters can retain their gas and dust for a longer period of time, and appear compact (in line emission) and dusty up to $\sim$6~Myr and possibly longer. The age limit for our analysis, $\sim$7~Myr, is discussed earlier  and is due to two concomitant effects: (1) beyond 7~Myr line emission becomes too faint to be recognized as such, and (2) there are age degeneracies in the SED fits for ages $>$6--7~Myr.

%We observe that the descendants of the massive clusters in the LEGUS sample do not retain their gas. 

We observe that the massive clusters in the LEGUS sample retain their gas only for the first $\sim$4~Myr, implying that timescales may be too short to observe  their dustier progenitors. Conversely, we speculate that the progenitors of the lower--mass, Pa$\beta$ emitting, compact clusters we identify are buried in dust and are undetectable in our near--IR bands, similar to what occurs in the Milky Way \citep{Gutermuth+2009,Megeath+2022}. This may account for why we do not find young and dusty star clusters in our sample. Alternatively, stochastic sampling of the IMF may affect our age determinations (see next sections), although we attempt to mitigate its impact by selecting clusters with mass $\ge$3,000~M$_{\odot}$.

\subsection{Comparison with NGC\,1313}

The study of \citet{Messa2021} of the late--type spiral NGC\,1313 is the closest one in methodology to the present study. These authors also find that the Pa$\beta$--selected young star clusters in NGC\,1313 are low--mass systems and show a plume of high extinctions concentrated around 4--10~Myr, and that the LEGUS star clusters for this galaxy identify typically lower--extinction sources. However, \citet{Messa2021} do not isolate the LEGUS clusters that are coincident with ionized gas emission and do not compare the LEGUS stellar masses with the stellar masses of their Pa$\beta$ selected sources.  Despite these differences, those authors conclude that when including infrared--selected star clusters the clearing timescales extend from $\sim$2~Myr to $\sim$4~Myr. Our timescales appear slightly longer, $\approx$6~Myr, for the Pa$\beta$--detected clusters. This difference, if confirmed by further studies, may be due to the higher pressure environment of NGC\,4449 relative to that of NGC\,1313. Based on the SFR surface density presented in the Introduction (section~\ref{sec:intro}), the environmental pressure in NGC\,4449 is about 2-3 times higher than in NGC\,1313.

\subsection{Consequences for the Timescale of Cluster Emergence}
Our relatively unbiased sample of Pa$\beta$--emitting compact sources in the galaxy NGC\,4449 shows a dearth of sources younger than 4~Myr, while, at the same time, we find that sources in the age range 5--6~Myr can display high values of dust extinction, up to E(B--V)$\sim 1.4$~mag. As discussed in section~\ref{sec:JWST}, none of the sources identified in \citet{Calzetti2023} with E(B$-$V)$\gtrsim$2~mag are confirmed by our more detailed inspection using the JWST Pa$\alpha$ image, although we still find, like those authors, that we do not recover very young ($\le$4~Myr) sources.  Conversely, the UV(U)--bright clusters identified in the LEGUS sample are associated with line emission only when $\lesssim$4~Myr old. 

As presented in Section \ref{sec:Source Catalog}, the original sample from SExtractor was visually inspected to remove stars, asterisms, image artifacts, and also sources where the Pa$\beta$ line emission is not coincident with the stellar continuum emission. This last criterion de facto removes from the sample sources where the ionized gas has been displaced from the location of the ionizing star cluster.  This can happen, for instance, if supernova or pre--supernova feedback has occurred in the cluster, indicating that our Pa$\beta$--emitting sample will be missing sources where feedback has been active.

Pre--SN feedback acts on short timescales ($<$4~Myr) before the occurrence of supernovae around $\sim$4~Myr 
%for clusters of our mass range ($>$3000 M$_\odot$) 
\citep{Pellegrini+2011,Dale+2012,Krause+2013,Krumholz2019, Leitherer+2014}. In star clusters where pre--SN feedback is efficient, the surrounding ionized gas is ejected, creating a `shell-like' morphology for the gas emission \citep{Whitmore+2011, Hollyhead+2015, Hannon+2022}. These clusters are likely to be young ($\lesssim$4~Myr), but also have low extinction, which explains why they have been prominently identified in UV--optical samples \citep{Messa2021, Hannon+2022}. 

When selecting for compact--emission sources in the near infrared, like our sample, we succeed in locating star clusters with a large range of dust extinction properties, including very dusty ones with E(B--V)$>$1~mag. However, this sample of compact, line--emitting sources lacks very young ($\lesssim$4~Myr) clusters, raising the question of how effective pre--SN and supernova feedback is in the $\sim$3,000--15,000~M$_{\odot}$ mass regime of our clusters. All of our clusters  contain massive stars (presence of ionized gas emission), yet they show compact morphology and include many dusty sources, suggesting that neither pre--SN feedback nor supernova feedback are effective clearing mechanisms in these sources. Even considering a range of supernova timescales, $\sim$4-8 Myr, due to incomplete sampling of the IMF \citep{Chevance+2022, Stanway+2023}, pre--SN feedback should still have performed a significant amount of clearing in these clusters.

Using the calculations presented in \citet{Calzetti2023} adapted to our range of extinctions (A$_V\sim$0.8--4.5~mag), we derive external pressures in the range P$_{ext}$/$k_B$=(0.1--2.3)$\times$10$^6$~K~cm$^{-3}$, which are significantly smaller, by about an order of magnitude, than the pressure supplied by photoionization, direct radiation and stellar winds from the star clusters \citep{Calzetti2023}. Thus, at face value, the host environment of our clusters does not represent an impediment to the clusters' ability to remove gas and dust from their immediate surroundings. 

However, the Pa$\beta$--emitting clusters, albeit selected to be $>$3,000~M$_{\odot}$, are a factor $\sim$7 lower mass than the LEGUS line--emitting clusters, with a median mass of 7,000~M$_{\odot}$. Therefore, we can expect stochastic sampling of the stellar IMF to have an impact on the cluster's properties, due to the effect this has on the number and mass range of the massive stars contained in the clusters \citep{Cervino2002, Stanway+2023}. \citet{Stanway+2023}, using the binary population models BPASS, find that a 7,000~M$_{\odot}$ cluster typically produces stars with maximum mass around 18~M$_{\odot}$, with a 16th-84th percentile range of $\sim$9--25~M$_{\odot}$. These are mid--O/early--B type stars, which, although still able to produce ionizing photons and stellar winds, generate rates that are an order of magnitude or more below those of a fully sampled IMF \citep{Leitherer+1999, Vink+2021}. 
%Most of these clusters' gas--clearing ability rests with supernovae \citep{Stanway+2023}. 
\citet{Chevance+2022} shows that, in models, stochastic sampling plays a role in the spread of expected delay times for supernovae to occur (see their Figure~4), covering the range $\sim$4-10~Myr for a 3,000~$M_{\odot}$ cluster, but settling at $\sim$4~Myr for clusters heavier than $\sim$2$\times$10$^4$~M$_{\odot}$. Stochastic sampling can also affect our age determinations, since we use deterministic stellar population models to derive our ages, masses and extinctions. However, deterministic models generally produce younger ages than stochastic models \citep{Krumholz+2015}, thus our clusters may well span a wider range of ages than those we derive, but will also tend to be older. 

Since external pressure in NGC\,4449 is about an order a magnitude too low to be a factor in determining the clearing timescale, the picture that emerges from these arguments is that pre--SN and supernova feedback are not effective on the same timescale in all star clusters, but will depend on the cluster's mass. The clearing timescale of 5--6~Myr we find is aligned with expectations from stochastic sampling of the IMF in the mass range of our line--emitting star clusters.

\section{Summary and Conclusions}\label{sec:summary}
We have utilized multi-wavelength HST photometry, including narrowband filters corresponding to H$\alpha$ and Pa$\beta$ hydrogen recombination lines, to build a sample of 99 young cluster candidates in the nearby dwarf-starburst NGC\,4449. Leveraging a JWST/NIRCam Pa$\alpha$ image as well as our own images, we visually confirmed that these sources possess compact morphologies with coincident line emission. We also found that none of the sources identified by \citet{Calzetti2023} shows line emission in the JWST Pa$\alpha$ image. After deriving the SEDs of the candidates using aperture photometry and fitting them to synthetic SED models, we were able to derive the ages, masses, and extinctions for all 99  candidates. We isolate a final catalog of 15 sources with $\chi_{red}^{2}$$<$10 and mass $>$3000 M$_{\odot}$ to minimize the effects of stochastic sampling of the stellar IMF. Our analysis of the resulting sources reveals a  notable plume of sources at 5-6 Myr. 
%The latter is adequately explained by a lack of strong Pa$\beta$ line emission exhibited by these sources which is required to break the degeneracy between the SEDs of young/dusty and old/dust--free reddened clusters. 
We also observe relatively few very young sources, younger than $\sim$4~Myr, and only one with mass$\ge$3,000~M$_{\odot}$. 
%Visual examination of the dust images of NGC\,4449 returned only one additional very young cluster, with modest color--excess. 
A critical evaluation of our selection method indicates that we are unlikely to have missed from our sample very young ($\lesssim$4~Myr) sources with compact line emission. 

A comparison with the UV--optically selected young star clusters from the LEGUS sample shows that, while our clusters retain their gas up to an age of at least 5--6~Myr, possibly longer, the LEGUS clusters lose their gas by 4~Myr age. The main difference between the two samples is that the young LEGUS clusters extend to masses much higher than those in our sample.  Considering only clusters with mass $\ge$3,000~M$_{\odot}$, the median mass for the LEGUS clusters is $\sim$3.5$\times$10$^4$~M$_{\odot}$, while for our sample it is $\sim$7$\times$10$^3$~M$_{\odot}$. These results suggest that pre--SN and SN feedback are  effective in massive star clusters, which, as a result, do not retain their gas (and dust) beyond $\sim$4~Myr. Conversely, they appear to be less effective and require longer timescales in the lower--mass star clusters (i.e., those in our sample), and these clusters still retain gas and dust up to the maximum age our SED fits can reliably determine ($\lesssim$7~Myr). We suggest that the progenitors of these lower--mass clusters are still sufficiently buried in their natal cloud to have escaped detection in our sample. Only after about 5--6~Myr do these dusty clusters begin to emerge from their natal clouds. 

The scenario implied by our data is that the clearing timescales due to pre--SN and supernova feedback are cluster mass--dependent; the timescales are not universally short but increase for decreasing star cluster mass. This is likely due to both pre--SN and supernova feedback being impacted by stochastic sampling of the stellar IMF, whose effect increases for decreasing mass. If correct, this scenario would have important repercussions on a galaxy's ability to leak ionizing photons, since the steep cluster mass function implies that galaxies contain many low--mass clusters and relatively few massive star clusters, at least at low redshift \citep[e.g.][]{Adamo+2017, Messa2018a, Cook+2019}. 
%contradicts current models of cluster emergence which describe a fast clearing timescale of $<$3 Myr. 
Observations of nearby galaxies with the JWST  possess improved sensitivity and angular resolution in the NIR and will likely provide an improved sample of star clusters covering the full range of ages, masses, and extinctions in the local universe, enabling accurate determinations of the clusters' clearing timescales.
%Further studies will thus be able to confirm or refute our current findings.
%We have excluded sources whose ionized gas emission is no longer coincident with the stellar emission, but these are also sources that are likely to have low extinction, because the mechanisms that expel gas also expel dust. 

%From the considerations above, we are led to conclude that very young and highly extincted sources are absent from our sample because they are still sufficiently buried in dust to be undetectable in our images. 

\begin{acknowledgments}

The authors thank the FEAST team (JWST Cycle 1 program \# 1783) for providing the stellar--continuum subtracted Pa$\alpha$ image of NGC\,4449.

Based on observations made with the NASA/ESA Hubble Space Telescope, obtained  at the Space Telescope Science Institute, which is operated by the Association of Universities for Research in Astronomy, Inc., under NASA contract NAS 5--26555. These observations are associated with program \# 15330. Support for program \# 15330 was provided by NASA through a grant from the Space Telescope Science Institute.

Based also on observations made with the NASA/ESA Hubble Space Telescope, and obtained from the Hubble Legacy Archive, 
which is a collaboration between the Space Telescope Science Institute (STScI/NASA), the Space Telescope European Coordinating Facility (ST-ECF/ESA) and the Canadian Astronomy Data Centre (CADC/NRC/CSA).

Most of the data presented in this paper were obtained from the Mikulski Archive for Space Telescopes (MAST)
at the Space Telescope Science Institute. The specific observations analyzed can be accessed via \dataset[https://doi.org/10.17909/8k2f-7205]{https://doi.org/10.17909/8k2f-7205}

This research has made use of the NASA/IPAC Extragalactic Database (NED) which is operated by the Jet
Propulsion Laboratory, California Institute of Technology, under contract with the National Aeronautics and Space
Administration.
\end{acknowledgments}

\bibliography{sample631}{}
\bibliographystyle{aasjournal}

\begin{longrotatetable}
\begin{deluxetable}{lrrrrrrrrrrrrr}
\centering 
\tablecolumns{12}
\tabletypesize{\tiny}
\tablewidth{0pt}
\tablecaption{Source Location and Photometry}
\tablehead{\colhead{(ID)} & \colhead{(RA(2000),DEC(2000))} & \colhead{(F275W)} & \colhead{(F336W)} & \colhead{(F435W)} & \colhead{(F550M)} & \colhead{(F555W)} & \colhead{(F658N)} & \colhead{(F814W)} & \colhead{(F110W)} & \colhead{(F128N)} & \colhead{(F160W)} \\\colhead{(1)} & \colhead{(2)} & \colhead{(3)} & \colhead{(4)} & \colhead{(5)} & \colhead{(6)} & \colhead{(7)} & \colhead{(8)} & \colhead{(9)} & \colhead{(10)} & \colhead{(11)} & \colhead{(12)}}
\label{tab:Source Table}
\startdata
1 & 12:28:12.74 +44:04:58.75 & 21.25$\pm$0.08 & 21.42$\pm$0.08 & 21.63$\pm$0.08 & 21.94$\pm$0.08 & 21.78$\pm$0.08 & 20.35$\pm$0.11 & 22.39$\pm$0.08 & 23.28$\pm$0.14 & 22.53$\pm$0.16 & 23.69$\pm$0.23\\
2 & 12:28:11.01 +44:05:02.02 & 23.49$\pm$0.12 & 23.56$\pm$0.11 & 23.15$\pm$0.1 & 22.54$\pm$0.1 & 22.68$\pm$0.1 & 21.59$\pm$0.12 & 21.51$\pm$0.09 & 21.53$\pm$0.15 & 20.99$\pm$0.15 & 21.22$\pm$0.16\\
3 & 12:28:08.85 +44:05:09.30 & 22.46$\pm$0.09 & 22.12$\pm$0.09 & 21.99$\pm$0.09 & 21.9$\pm$0.09 & 21.53$\pm$0.08 & 18.7$\pm$0.09 & 21.31$\pm$0.08 & 21.63$\pm$0.1 & 20.09$\pm$0.1 & 21.81$\pm$0.11\\
4 & 12:28:13.86 +44:05:12.03 & 20.72$\pm$0.08 & 20.9$\pm$0.08 & 20.75$\pm$0.08 & 21.08$\pm$0.08 & 20.98$\pm$0.08 & 20.05$\pm$0.09 & 21.35$\pm$0.08 & 22.79$\pm$0.16 & 22.16$\pm$0.12 & 23.22$\pm$0.16\\
5 & 12:28:08.95 +44:05:13.16 & 25.96$\pm$1.09 & 26.36$\pm$1.45 & 25.34$\pm$0.84 & 23.29$\pm$0.17 & 23.55$\pm$0.21 & 22.52$\pm$0.33 & 22.25$\pm$0.1 & 22.27$\pm$0.18 & 21.98$\pm$0.18 & 21.95$\pm$0.21\\
6 & 12:28:08.77 +44:05:12.83 & 25.06$\pm$0.38 & 25.38$\pm$0.49 & 25.22$\pm$0.68 & 23.65$\pm$0.17 & 23.52$\pm$0.2 & 20.49$\pm$0.16 & 22.54$\pm$0.1 & 22.95$\pm$0.21 & 22.25$\pm$0.18 & 22.58$\pm$0.16\\
7 & 12:28:09.29 +44:05:15.05 & 20.56$\pm$0.08 & 20.69$\pm$0.08 & 20.9$\pm$0.08 & 21.13$\pm$0.08 & 21.14$\pm$0.08 & 19.12$\pm$0.09 & 21.67$\pm$0.09 & 22.93$\pm$0.25 & 21.88$\pm$0.2 & 23.53$\pm$0.45\\
8 & 12:28:09.39 +44:05:15.30 & 24.26$\pm$0.59 & 26.43$\pm$5.55 & 23.77$\pm$0.3 & 22.3$\pm$0.1 & 22.76$\pm$0.16 & 22.21$\pm$1.09 & 21.42$\pm$0.08 & 21.71$\pm$0.2 & 21.28$\pm$0.18 & 21.39$\pm$0.14\\
9 & 12:28:09.95 +44:05:15.87 & 23.12$\pm$0.14 & 23.27$\pm$0.18 & 23.33$\pm$0.22 & 23.51$\pm$0.27 & 23.2$\pm$0.22 & 21.17$\pm$0.24 & 22.16$\pm$0.12 & 21.95$\pm$0.34 & 21.47$\pm$0.21 & 21.39$\pm$0.32\\
10 & 12:28:10.69 +44:05:17.36 & 28.54$\pm$4.58 & 24.8$\pm$0.2 & 23.29$\pm$0.1 & 22.68$\pm$0.1 & 22.63$\pm$0.09 & 22.0$\pm$0.29 & 22.04$\pm$0.1 & 22.39$\pm$0.28 & 22.01$\pm$0.3 & 22.22$\pm$0.39\\
11 & 12:28:09.32 +44:05:17.01 & 20.82$\pm$0.08 & 20.94$\pm$0.08 & 21.15$\pm$0.08 & 21.49$\pm$0.08 & 21.35$\pm$0.08 & 19.07$\pm$0.08 & 21.8$\pm$0.08 & 22.84$\pm$0.27 & 21.66$\pm$0.14 & 23.5$\pm$0.82\\
12 & 12:28:09.71 +44:05:20.59 & 22.99$\pm$0.12 & 22.78$\pm$0.1 & 22.29$\pm$0.11 & 21.47$\pm$0.1 & 21.72$\pm$0.1 & 21.6$\pm$0.24 & 20.43$\pm$0.08 & 20.55$\pm$0.14 & 20.12$\pm$0.15 & 20.24$\pm$0.13\\
13 & 12:28:09.53 +44:05:21.01 & 25.86$\pm$0.8 & 25.63$\pm$0.62 & 26.44$\pm$1.31 & 23.7$\pm$0.14 & 23.91$\pm$0.16 & 21.97$\pm$0.44 & 21.91$\pm$0.09 & 21.78$\pm$0.12 & 21.09$\pm$0.12 & 21.33$\pm$0.12\\
14 & 12:28:08.91 +44:05:22.52 & 22.84$\pm$0.09 & 22.61$\pm$0.08 & 22.57$\pm$0.09 & 22.66$\pm$0.09 & 22.27$\pm$0.08 & 19.69$\pm$0.08 & 22.44$\pm$0.09 & 23.08$\pm$0.11 & 21.76$\pm$0.12 & 23.76$\pm$0.29\\
15 & 12:28:13.04 +44:05:22.83 & 21.55$\pm$0.08 & 21.57$\pm$0.08 & 21.84$\pm$0.08 & 21.88$\pm$0.08 & 22.07$\pm$0.08 & 22.1$\pm$0.2 & 22.19$\pm$0.08 & 23.14$\pm$0.12 & 22.72$\pm$0.22 & 23.36$\pm$0.23\\
16 & 12:28:12.37 +44:05:23.19 & 19.42$\pm$0.08 & 19.46$\pm$0.08 & 19.56$\pm$0.08 & 19.74$\pm$0.08 & 20.04$\pm$0.08 & 18.74$\pm$0.08 & 20.12$\pm$0.08 & 20.98$\pm$0.11 & 20.2$\pm$0.1 & 21.37$\pm$0.14\\
17 & 12:28:10.08 +44:05:23.52 & 25.78$\pm$1.64 & 24.23$\pm$0.39 & 23.83$\pm$0.31 & 23.53$\pm$0.22 & 23.36$\pm$0.22 & 21.83$\pm$0.53 & 22.54$\pm$0.11 & 22.62$\pm$0.2 & 22.03$\pm$0.24 & 22.26$\pm$0.22\\
18 & 12:28:09.99 +44:05:24.66 & 20.71$\pm$0.09 & 20.53$\pm$0.09 & 20.46$\pm$0.09 & 20.64$\pm$0.09 & 19.81$\pm$0.09 & 17.58$\pm$0.1 & 20.45$\pm$0.09 & 20.94$\pm$0.15 & 19.46$\pm$0.13 & 21.23$\pm$0.14\\
19 & 12:28:13.94 +44:05:30.56 & 25.5$\pm$0.25 & 25.1$\pm$0.14 & 24.65$\pm$0.12 & 24.21$\pm$0.14 & 23.86$\pm$0.1 & 21.25$\pm$0.09 & 23.53$\pm$0.11 & 23.82$\pm$0.28 & 22.52$\pm$0.14 & 23.92$\pm$0.48\\
20 & 12:28:11.36 +44:05:30.90 & 26.35$\pm$1.09 & 26.07$\pm$1.09 & 25.47$\pm$1.09 & 26.07$\pm$2.0 & 25.6$\pm$1.09 & 22.26$\pm$0.29 & 21.96$\pm$0.11 & 21.82$\pm$0.16 & 21.15$\pm$0.13 & 21.26$\pm$0.15\\
21 & 12:28:10.43 +44:05:30.19 & 20.22$\pm$0.08 & 20.1$\pm$0.08 & 19.89$\pm$0.08 & 19.55$\pm$0.08 & 19.64$\pm$0.08 & 18.88$\pm$0.08 & 18.81$\pm$0.08 & 19.0$\pm$0.12 & 18.52$\pm$0.11 & 18.72$\pm$0.12\\
22 & 12:28:10.67 +44:05:32.23 & 24.26$\pm$0.18 & 24.05$\pm$0.18 & 23.8$\pm$0.23 & 23.13$\pm$0.18 & 22.97$\pm$0.16 & 20.79$\pm$0.14 & 21.14$\pm$0.09 & 20.84$\pm$0.17 & 20.12$\pm$0.15 & 20.23$\pm$0.17\\
23 & 12:28:09.38 +44:05:35.08 & 21.98$\pm$0.09 & 22.16$\pm$0.09 & 22.21$\pm$0.09 & 21.81$\pm$0.08 & 21.92$\pm$0.09 & 21.07$\pm$0.1 & 20.98$\pm$0.08 & 21.17$\pm$0.11 & 20.71$\pm$0.1 & 20.82$\pm$0.11\\
24 & 12:28:11.51 +44:05:37.62 & 26.03$\pm$1.09 & 26.01$\pm$1.09 & 25.07$\pm$1.09 & 22.94$\pm$0.24 & 23.22$\pm$0.29 & 21.98$\pm$0.28 & 20.6$\pm$0.08 & 20.44$\pm$0.1 & 19.91$\pm$0.1 & 20.16$\pm$0.1\\
25 & 12:28:10.80 +44:05:36.80 & 20.35$\pm$0.08 & 20.28$\pm$0.08 & 20.22$\pm$0.08 & 20.41$\pm$0.08 & 19.98$\pm$0.09 & 17.96$\pm$0.09 & 20.57$\pm$0.08 & 21.38$\pm$0.21 & 19.92$\pm$0.12 & 22.04$\pm$0.52\\
26 & 12:28:11.40 +44:05:37.21 & 23.01$\pm$0.15 & 22.59$\pm$0.13 & 22.7$\pm$0.2 & 22.43$\pm$0.17 & 21.92$\pm$0.19 & 19.37$\pm$0.14 & 21.08$\pm$0.09 & 20.85$\pm$0.13 & 20.16$\pm$0.12 & 20.28$\pm$0.13\\
27 & 12:28:17.87 +44:05:37.69 & 21.82$\pm$0.08 & 21.92$\pm$0.08 & 21.9$\pm$0.08 & 22.06$\pm$0.08 & 21.9$\pm$0.1 & 20.55$\pm$0.2 & 22.37$\pm$0.09 & 23.1$\pm$0.15 & 21.95$\pm$0.15 & 23.32$\pm$0.14\\
28 & 12:28:12.25 +44:05:38.92 & 21.19$\pm$0.09 & 20.68$\pm$0.08 & 20.11$\pm$0.08 & 20.2$\pm$0.08 & 20.13$\pm$0.08 & 19.41$\pm$0.08 & 20.3$\pm$0.08 & 21.12$\pm$0.14 & 20.78$\pm$0.13 & 21.19$\pm$0.14\\
29 & 12:28:11.13 +44:05:40.50 & 24.82$\pm$0.32 & 24.73$\pm$0.52 & 23.34$\pm$0.22 & 21.36$\pm$0.09 & 21.64$\pm$0.1 & 20.31$\pm$0.13 & 19.92$\pm$0.08 & 19.95$\pm$0.12 & 19.41$\pm$0.11 & 19.63$\pm$0.13\\
30 & 12:28:10.93 +44:05:40.63 & 20.42$\pm$0.08 & 19.2$\pm$0.08 & 18.78$\pm$0.08 & 18.59$\pm$0.08 & 18.61$\pm$0.08 & 16.76$\pm$0.08 & 18.32$\pm$0.08 & 19.1$\pm$0.1 & 18.48$\pm$0.09 & 19.22$\pm$0.09\\
31 & 12:28:11.31 +44:05:42.29 & 21.73$\pm$0.09 & 21.79$\pm$0.09 & 21.98$\pm$0.13 & 22.38$\pm$0.16 & 22.11$\pm$0.14 & 21.07$\pm$0.12 & 22.21$\pm$0.15 & 22.51$\pm$0.36 & 22.01$\pm$0.26 & 22.31$\pm$0.38\\
32 & 12:28:13.09 +44:05:42.65 & 21.24$\pm$0.1 & 20.85$\pm$0.09 & 21.0$\pm$0.09 & 21.42$\pm$0.09 & 19.57$\pm$0.08 & 17.03$\pm$0.08 & 20.58$\pm$0.08 & 20.98$\pm$0.14 & 19.2$\pm$0.11 & 21.5$\pm$0.14\\
33 & 12:28:11.16 +44:05:41.33 & 25.1$\pm$0.4 & 24.89$\pm$0.3 & 23.5$\pm$0.17 & 22.21$\pm$0.11 & 22.41$\pm$0.12 & 21.3$\pm$0.15 & 20.71$\pm$0.09 & 20.53$\pm$0.14 & 20.01$\pm$0.13 & 20.14$\pm$0.14\\
34 & 12:28:12.18 +44:05:42.61 & 21.36$\pm$0.1 & 21.14$\pm$0.1 & 20.55$\pm$0.1 & 20.24$\pm$0.1 & 20.26$\pm$0.1 & 19.8$\pm$0.1 & 19.32$\pm$0.08 & 19.27$\pm$0.11 & 18.7$\pm$0.1 & 18.85$\pm$0.1\\
35 & 12:28:12.89 +44:05:44.24 & 20.87$\pm$0.09 & 20.97$\pm$0.09 & 21.09$\pm$0.1 & 21.45$\pm$0.1 & 21.13$\pm$0.1 & 19.29$\pm$0.09 & 21.51$\pm$0.1 & 22.02$\pm$0.15 & 21.2$\pm$0.16 & 21.9$\pm$0.16\\
36 & 12:28:13.08 +44:05:43.88 & 20.18$\pm$0.08 & 20.18$\pm$0.08 & 20.19$\pm$0.08 & 20.38$\pm$0.08 & 20.25$\pm$0.09 & 19.63$\pm$0.2 & 20.79$\pm$0.08 & 21.86$\pm$0.19 & 20.92$\pm$0.22 & 22.36$\pm$0.35\\
37 & 12:28:13.69 +44:05:44.76 & 19.25$\pm$0.08 & 19.49$\pm$0.08 & 19.66$\pm$0.08 & 20.07$\pm$0.08 & 19.94$\pm$0.08 & 19.85$\pm$0.12 & 20.79$\pm$0.08 & 21.93$\pm$0.13 & 21.63$\pm$0.16 & 22.48$\pm$0.15\\
38 & 12:28:11.94 +44:05:50.12 & 21.62$\pm$0.08 & 21.75$\pm$0.08 & 21.9$\pm$0.09 & 21.38$\pm$0.09 & 21.43$\pm$0.09 & 20.51$\pm$0.11 & 20.24$\pm$0.08 & 20.21$\pm$0.14 & 19.73$\pm$0.13 & 19.84$\pm$0.13\\
39 & 12:28:11.81 +44:05:52.17 & 24.41$\pm$0.18 & 24.91$\pm$0.35 & 24.6$\pm$0.33 & 25.67$\pm$1.22 & 24.35$\pm$0.36 & 22.55$\pm$0.58 & 22.06$\pm$0.09 & 21.7$\pm$0.12 & 21.1$\pm$0.11 & 21.2$\pm$0.11\\
40 & 12:28:11.70 +44:05:51.00 & 19.12$\pm$0.08 & 18.92$\pm$0.08 & 19.41$\pm$0.08 & 19.44$\pm$0.08 & 19.36$\pm$0.08 & 17.02$\pm$0.08 & 19.17$\pm$0.08 & 20.04$\pm$0.1 & 19.12$\pm$0.09 & 20.1$\pm$0.1\\
41 & 12:28:12.03 +44:05:52.65 & 26.58$\pm$4.82 & 24.85$\pm$1.09 & 25.79$\pm$3.04 & 23.3$\pm$0.24 & 23.2$\pm$0.34 & 20.93$\pm$0.42 & 20.36$\pm$0.08 & 19.94$\pm$0.12 & 19.37$\pm$0.11 & 19.54$\pm$0.13\\
42 & 12:28:11.82 +44:05:53.83 & 21.67$\pm$0.09 & 21.8$\pm$0.09 & 22.09$\pm$0.1 & 22.53$\pm$0.11 & 21.97$\pm$0.09 & 19.64$\pm$0.1 & 22.56$\pm$0.15 & 23.26$\pm$0.49 & 21.91$\pm$0.25 & 23.42$\pm$0.75\\
43 & 12:28:11.38 +44:05:53.26 & 24.76$\pm$0.17 & 24.9$\pm$0.18 & 24.03$\pm$0.13 & 23.21$\pm$0.11 & 23.35$\pm$0.11 & 22.45$\pm$0.18 & 21.93$\pm$0.08 & 21.51$\pm$0.1 & 20.91$\pm$0.1 & 21.12$\pm$0.09\\
44 & 12:28:12.03 +44:05:53.29 & 24.43$\pm$0.64 & 24.79$\pm$1.12 & 23.19$\pm$0.29 & 21.35$\pm$0.1 & 21.57$\pm$0.11 & 20.19$\pm$0.14 & 19.88$\pm$0.08 & 19.94$\pm$0.12 & 19.43$\pm$0.11 & 19.52$\pm$0.11\\
45 & 12:28:12.45 +44:05:54.83 & 26.69$\pm$1.48 & 26.43$\pm$1.33 & 27.18$\pm$4.37 & 23.76$\pm$0.21 & 23.98$\pm$0.26 & 22.85$\pm$0.41 & 21.55$\pm$0.08 & 21.26$\pm$0.12 & 20.74$\pm$0.12 & 20.76$\pm$0.12\\
46 & 12:28:19.11 +44:06:00.88 & 22.26$\pm$0.08 & 22.32$\pm$0.08 & 22.46$\pm$0.08 & 22.76$\pm$0.08 & 22.26$\pm$0.08 & 20.12$\pm$0.1 & 22.94$\pm$0.09 & 23.74$\pm$0.16 & 22.42$\pm$0.15 & 24.26$\pm$0.23\\
47 & 12:28:13.79 +44:06:02.82 & 22.02$\pm$0.08 & 21.54$\pm$0.08 & 21.44$\pm$0.08 & 21.44$\pm$0.08 & 21.38$\pm$0.08 & 19.91$\pm$0.08 & 21.33$\pm$0.08 & 22.31$\pm$0.18 & 21.64$\pm$0.13 & 22.26$\pm$0.24\\
48 & 12:28:18.60 +44:06:05.50 & 21.85$\pm$0.08 & 21.92$\pm$0.08 & 22.12$\pm$0.08 & 22.36$\pm$0.08 & 21.92$\pm$0.08 & 19.75$\pm$0.09 & 22.59$\pm$0.08 & 23.73$\pm$0.19 & 22.28$\pm$0.16 & 24.67$\pm$0.49\\
49 & 12:28:12.96 +44:06:06.54 & 26.24$\pm$0.73 & 26.18$\pm$0.81 & 23.97$\pm$0.17 & 22.22$\pm$0.09 & 22.52$\pm$0.09 & 21.28$\pm$0.12 & 20.89$\pm$0.08 & 20.82$\pm$0.11 & 20.4$\pm$0.1 & 20.45$\pm$0.11\\
50 & 12:28:14.00 +44:06:12.67 & 25.16$\pm$0.22 & 25.41$\pm$0.35 & 24.11$\pm$0.15 & 22.89$\pm$0.1 & 22.95$\pm$0.13 & 21.79$\pm$0.31 & 22.1$\pm$0.09 & 22.45$\pm$0.22 & 21.81$\pm$0.2 & 22.3$\pm$0.23\\
51 & 12:28:15.02 +44:06:18.90 & 22.03$\pm$0.08 & 21.76$\pm$0.08 & 21.66$\pm$0.08 & 21.51$\pm$0.08 & 21.48$\pm$0.08 & 19.4$\pm$0.08 & 21.4$\pm$0.08 & 22.16$\pm$0.12 & 21.29$\pm$0.11 & 22.22$\pm$0.15\\
52 & 12:28:10.63 +44:06:20.64 & 20.64$\pm$0.08 & 20.49$\pm$0.08 & 20.86$\pm$0.08 & 20.84$\pm$0.08 & 20.79$\pm$0.08 & 18.59$\pm$0.08 & 20.71$\pm$0.08 & 21.6$\pm$0.14 & 20.9$\pm$0.1 & 21.45$\pm$0.15\\
53 & 12:28:16.00 +44:06:29.31 & 20.2$\pm$0.08 & 19.87$\pm$0.08 & 19.82$\pm$0.08 & 20.15$\pm$0.08 & 18.29$\pm$0.08 & 16.0$\pm$0.08 & 19.5$\pm$0.08 & 19.84$\pm$0.11 & 18.24$\pm$0.1 & 20.37$\pm$0.11\\
54 & 12:28:15.75 +44:06:27.98 & 25.19$\pm$0.17 & 24.46$\pm$0.11 & 24.34$\pm$0.1 & 24.44$\pm$0.16 & 23.78$\pm$0.1 & 20.86$\pm$0.12 & 23.74$\pm$0.1 & 24.27$\pm$0.36 & 22.58$\pm$0.2 & 23.94$\pm$0.4\\
55 & 12:28:15.87 +44:06:29.97 & 21.36$\pm$0.08 & 21.38$\pm$0.08 & 21.37$\pm$0.08 & 21.45$\pm$0.08 & 21.44$\pm$0.09 & 21.56$\pm$0.36 & 21.72$\pm$0.08 & 22.82$\pm$0.15 & 22.48$\pm$0.27 & 23.36$\pm$0.17\\
56 & 12:28:16.35 +44:06:32.06 & 22.8$\pm$0.09 & 22.87$\pm$0.09 & 23.21$\pm$0.09 & 23.4$\pm$0.1 & 23.07$\pm$0.09 & 21.08$\pm$0.12 & 23.78$\pm$0.11 & 24.97$\pm$0.44 & 23.52$\pm$0.23 & 27.32$\pm$5.02\\
57 & 12:28:10.23 +44:06:35.85 & 20.34$\pm$0.08 & 20.32$\pm$0.08 & 19.7$\pm$0.08 & 19.78$\pm$0.08 & 19.72$\pm$0.08 & 18.62$\pm$0.08 & 20.1$\pm$0.08 & 21.1$\pm$0.1 & 20.43$\pm$0.09 & 21.43$\pm$0.09\\
58 & 12:28:10.86 +44:06:43.57 & 21.78$\pm$0.08 & 21.84$\pm$0.08 & 22.0$\pm$0.08 & 22.38$\pm$0.08 & 21.82$\pm$0.08 & 19.55$\pm$0.08 & 22.53$\pm$0.08 & 23.27$\pm$0.11 & 22.04$\pm$0.11 & 23.7$\pm$0.12\\
59 & 12:28:15.11 +44:06:45.30 & 24.3$\pm$0.14 & 23.87$\pm$0.09 & 23.62$\pm$0.09 & 23.9$\pm$0.15 & 22.34$\pm$0.08 & 20.03$\pm$0.08 & 23.12$\pm$0.09 & 23.26$\pm$0.12 & 21.89$\pm$0.11 & 23.54$\pm$0.16\\
60 & 12:28:12.25 +44:06:46.88 & 22.22$\pm$0.08 & 22.4$\pm$0.08 & 22.49$\pm$0.08 & 22.93$\pm$0.09 & 22.63$\pm$0.09 & 21.78$\pm$0.19 & 23.28$\pm$0.09 & 24.48$\pm$0.21 & 24.09$\pm$0.46 & 24.85$\pm$0.36\\
61 & 12:28:13.90 +44:06:49.11 & 20.52$\pm$0.08 & 20.75$\pm$0.08 & 20.86$\pm$0.08 & 21.16$\pm$0.08 & 21.06$\pm$0.09 & 20.45$\pm$0.2 & 21.53$\pm$0.08 & 22.58$\pm$0.15 & 22.08$\pm$0.25 & 22.92$\pm$0.18\\
62 & 12:28:13.05 +44:06:49.87 & 22.04$\pm$0.08 & 21.75$\pm$0.08 & 21.61$\pm$0.08 & 21.56$\pm$0.08 & 21.62$\pm$0.08 & 20.18$\pm$0.08 & 21.95$\pm$0.08 & 22.53$\pm$0.14 & 21.55$\pm$0.1 & 22.49$\pm$0.19\\
63 & 12:28:15.92 +44:06:29.60 & 25.24$\pm$0.32 & 24.83$\pm$0.25 & 24.56$\pm$0.21 & 24.52$\pm$0.25 & 24.39$\pm$0.21 & 21.11$\pm$0.17 & 24.38$\pm$0.22 & 24.54$\pm$0.37 & 22.77$\pm$0.17 & 24.68$\pm$0.36\\
64 & 12:28:12.95 +44:04:59.27 & 19.11$\pm$0.08 & 19.3$\pm$0.08 & 19.57$\pm$0.08 & 19.92$\pm$0.08 & 19.86$\pm$0.09 & 19.17$\pm$0.25 & 20.41$\pm$0.08 & 21.48$\pm$0.17 & 20.97$\pm$0.34 & 21.84$\pm$0.17\\
65 & 12:28:12.65 +44:04:58.85 & 24.9$\pm$0.36 & 24.09$\pm$0.22 & 24.13$\pm$0.21 & 23.94$\pm$0.14 & 23.16$\pm$0.13 & 20.28$\pm$0.11 & 23.19$\pm$0.1 & 23.42$\pm$0.24 & 21.88$\pm$0.14 & 23.36$\pm$0.15\\
66 & 12:28:12.56 +44:05:00.26 & 23.08$\pm$0.09 & 22.87$\pm$0.09 & 22.82$\pm$0.09 & 23.28$\pm$0.1 & 22.25$\pm$0.09 & 19.6$\pm$0.09 & 23.02$\pm$0.11 & 23.42$\pm$0.23 & 22.09$\pm$0.18 & 23.25$\pm$0.22\\
67 & 12:28:14.47 +44:05:01.79 & 22.14$\pm$0.08 & 22.2$\pm$0.09 & 22.27$\pm$0.09 & 22.68$\pm$0.09 & 22.05$\pm$0.08 & 19.67$\pm$0.08 & 22.74$\pm$0.09 & 23.41$\pm$0.2 & 22.2$\pm$0.12 & 23.52$\pm$0.33\\
68 & 12:28:09.06 +44:05:10.49 & 24.51$\pm$0.26 & 24.08$\pm$0.16 & 23.27$\pm$0.12 & 23.31$\pm$0.19 & 22.65$\pm$0.11 & 20.64$\pm$0.15 & 22.12$\pm$0.1 & 21.88$\pm$0.13 & 21.22$\pm$0.12 & 21.44$\pm$0.13\\
69 & 12:28:09.17 +44:05:12.32 & 20.86$\pm$0.09 & 21.12$\pm$0.09 & 21.32$\pm$0.09 & 21.84$\pm$0.1 & 21.63$\pm$0.1 & 20.05$\pm$0.1 & 22.07$\pm$0.1 & 22.7$\pm$0.2 & 22.04$\pm$0.17 & 22.75$\pm$0.24\\
70 & 12:28:14.84 +44:05:13.00 & 20.09$\pm$0.08 & 20.36$\pm$0.08 & 20.53$\pm$0.08 & 20.88$\pm$0.08 & 20.77$\pm$0.09 & 20.77$\pm$0.43 & 21.41$\pm$0.08 & 22.48$\pm$0.16 & 21.94$\pm$0.36 & 22.71$\pm$0.16\\
71 & 12:28:10.53 +44:05:15.90 & 25.91$\pm$1.09 & 26.28$\pm$1.09 & 25.78$\pm$1.09 & 24.14$\pm$0.37 & 24.6$\pm$0.5 & 22.46$\pm$0.27 & 21.75$\pm$0.1 & 21.38$\pm$0.11 & 20.83$\pm$0.11 & 20.91$\pm$0.11\\
72 & 12:28:09.06 +44:05:15.74 & 22.4$\pm$0.09 & 22.33$\pm$0.09 & 22.23$\pm$0.09 & 22.19$\pm$0.09 & 22.03$\pm$0.09 & 19.97$\pm$0.11 & 21.93$\pm$0.08 & 22.67$\pm$0.17 & 21.58$\pm$0.13 & 22.58$\pm$0.24\\
73 & 12:28:09.41 +44:05:16.02 & 21.67$\pm$0.09 & 21.66$\pm$0.09 & 21.8$\pm$0.09 & 22.13$\pm$0.1 & 21.73$\pm$0.1 & 18.94$\pm$0.09 & 22.06$\pm$0.1 & 22.93$\pm$0.3 & 21.35$\pm$0.17 & 23.2$\pm$0.61\\
74 & 12:28:09.22 +44:05:17.63 & 21.58$\pm$0.09 & 21.54$\pm$0.09 & 21.77$\pm$0.08 & 22.2$\pm$0.09 & 21.13$\pm$0.08 & 18.47$\pm$0.09 & 21.8$\pm$0.09 & 22.15$\pm$0.17 & 20.71$\pm$0.14 & 22.54$\pm$0.3\\
75 & 12:28:09.34 +44:05:17.99 & 24.08$\pm$0.15 & 24.01$\pm$0.12 & 23.8$\pm$0.11 & 23.77$\pm$0.16 & 23.46$\pm$0.12 & 20.69$\pm$0.09 & 23.35$\pm$0.14 & 25.1$\pm$2.27 & 22.36$\pm$0.2 & 23.68$\pm$1.09\\
76 & 12:28:12.92 +44:05:20.08 & 22.53$\pm$0.08 & 22.48$\pm$0.09 & 22.4$\pm$0.08 & 22.2$\pm$0.08 & 22.58$\pm$0.11 & 22.89$\pm$1.09 & 22.79$\pm$0.09 & 23.79$\pm$0.22 & 24.06$\pm$0.85 & 24.0$\pm$0.29\\
77 & 12:28:08.95 +44:05:22.19 & 25.3$\pm$0.3 & 24.68$\pm$0.19 & 24.13$\pm$0.17 & 24.18$\pm$0.18 & 23.82$\pm$0.14 & 20.96$\pm$0.1 & 23.94$\pm$0.14 & 24.86$\pm$0.91 & 22.51$\pm$0.24 & 24.85$\pm$2.0\\
78 & 12:28:13.61 +44:05:24.95 & 22.37$\pm$0.08 & 22.3$\pm$0.08 & 22.22$\pm$0.08 & 22.29$\pm$0.08 & 22.25$\pm$0.08 & 21.74$\pm$0.19 & 22.5$\pm$0.08 & 23.48$\pm$0.15 & 23.07$\pm$0.19 & 24.15$\pm$0.33\\
79 & 12:28:13.65 +44:05:28.32 & 22.76$\pm$0.1 & 22.43$\pm$0.1 & 22.5$\pm$0.1 & 22.73$\pm$0.1 & 21.73$\pm$0.11 & 19.34$\pm$0.11 & 22.59$\pm$0.12 & 22.95$\pm$0.34 & 21.42$\pm$0.24 & 23.76$\pm$1.0\\
80 & 12:28:13.72 +44:05:27.72 & 25.18$\pm$0.2 & 24.42$\pm$0.12 & 24.26$\pm$0.11 & 24.08$\pm$0.12 & 23.52$\pm$0.1 & 20.72$\pm$0.1 & 23.38$\pm$0.1 & 23.48$\pm$0.26 & 22.08$\pm$0.17 & 23.2$\pm$0.39\\
81 & 12:28:10.90 +44:05:32.62 & 19.83$\pm$0.08 & 19.78$\pm$0.08 & 19.55$\pm$0.09 & 19.56$\pm$0.09 & 19.55$\pm$0.09 & 19.49$\pm$0.1 & 19.05$\pm$0.08 & 19.1$\pm$0.18 & 18.67$\pm$0.19 & 18.72$\pm$0.2\\
82 & 12:28:11.11 +44:05:38.21 & 23.0$\pm$0.68 & 22.58$\pm$0.47 & 22.1$\pm$0.4 & 21.23$\pm$0.19 & 21.32$\pm$0.24 & 19.34$\pm$0.21 & 19.73$\pm$0.11 & 19.82$\pm$0.27 & 19.07$\pm$0.19 & 19.4$\pm$0.26\\
83 & 12:28:12.76 +44:05:40.71 & 21.45$\pm$0.08 & 21.6$\pm$0.08 & 21.56$\pm$0.08 & 21.92$\pm$0.09 & 21.67$\pm$0.08 & 19.66$\pm$0.09 & 22.22$\pm$0.11 & 23.13$\pm$0.38 & 22.13$\pm$0.26 & 23.4$\pm$0.73\\
84 & 12:28:11.84 +44:05:41.03 & 22.8$\pm$0.12 & 22.41$\pm$0.11 & 22.35$\pm$0.11 & 22.64$\pm$0.12 & 21.15$\pm$0.09 & 18.28$\pm$0.09 & 21.75$\pm$0.09 & 21.9$\pm$0.19 & 19.94$\pm$0.14 & 22.24$\pm$0.3\\
85 & 12:28:18.20 +44:05:46.70 & 21.84$\pm$0.08 & 21.97$\pm$0.08 & 22.09$\pm$0.08 & 22.28$\pm$0.08 & 22.01$\pm$0.09 & 20.42$\pm$0.1 & 22.6$\pm$0.08 & 23.44$\pm$0.15 & 22.27$\pm$0.14 & 23.77$\pm$0.2\\
86 & 12:28:11.96 +44:05:47.30 & 19.57$\pm$0.09 & 19.44$\pm$0.09 & 18.9$\pm$0.08 & 18.83$\pm$0.08 & 18.83$\pm$0.08 & 18.66$\pm$0.09 & 18.57$\pm$0.08 & 18.8$\pm$0.12 & 18.45$\pm$0.12 & 18.5$\pm$0.14\\
87 & 12:28:19.20 +44:06:00.64 & 22.81$\pm$0.09 & 22.46$\pm$0.08 & 22.69$\pm$0.09 & 22.91$\pm$0.09 & 22.3$\pm$0.09 & 20.08$\pm$0.1 & 23.02$\pm$0.09 & 23.59$\pm$0.18 & 22.45$\pm$0.13 & 23.75$\pm$0.36\\
88 & 12:28:13.57 +44:06:12.32 & 21.24$\pm$0.08 & 21.41$\pm$0.08 & 21.46$\pm$0.08 & 21.92$\pm$0.09 & 21.72$\pm$0.08 & 20.06$\pm$0.1 & 22.25$\pm$0.1 & 23.1$\pm$0.29 & 22.31$\pm$0.27 & 23.44$\pm$0.54\\
89 & 12:28:16.48 +44:06:22.64 & 23.92$\pm$0.1 & 23.45$\pm$0.09 & 23.4$\pm$0.09 & 23.59$\pm$0.1 & 22.52$\pm$0.08 & 19.73$\pm$0.08 & 23.02$\pm$0.09 & 23.58$\pm$0.27 & 21.95$\pm$0.14 & 24.07$\pm$0.43\\
90 & 12:28:16.02 +44:06:27.82 & 22.93$\pm$0.09 & 22.88$\pm$0.09 & 23.04$\pm$0.08 & 23.45$\pm$0.1 & 22.66$\pm$0.08 & 20.16$\pm$0.08 & 23.3$\pm$0.1 & 23.71$\pm$0.3 & 22.67$\pm$0.22 & 24.08$\pm$0.64\\
91 & 12:28:11.62 +44:06:31.15 & 22.04$\pm$0.08 & 22.08$\pm$0.08 & 22.25$\pm$0.08 & 22.67$\pm$0.09 & 22.15$\pm$0.09 & 19.86$\pm$0.11 & 22.7$\pm$0.09 & 23.66$\pm$0.31 & 22.15$\pm$0.17 & 23.96$\pm$0.69\\
92 & 12:28:14.06 +44:06:30.21 & 23.05$\pm$0.09 & 23.03$\pm$0.09 & 23.3$\pm$0.1 & 23.53$\pm$0.11 & 23.28$\pm$0.1 & 21.1$\pm$0.12 & 23.78$\pm$0.11 & 24.72$\pm$0.3 & 23.14$\pm$0.22 & 25.58$\pm$0.38\\
93 & 12:28:11.38 +44:06:31.43 & 21.31$\pm$0.08 & 21.11$\pm$0.08 & 21.13$\pm$0.08 & 20.9$\pm$0.08 & 20.98$\pm$0.08 & 19.47$\pm$0.08 & 20.6$\pm$0.08 & 20.99$\pm$0.11 & 20.39$\pm$0.1 & 20.73$\pm$0.12\\
94 & 12:28:11.63 +44:06:32.33 & 22.51$\pm$0.09 & 22.54$\pm$0.09 & 22.55$\pm$0.09 & 22.84$\pm$0.1 & 22.57$\pm$0.1 & 20.45$\pm$0.1 & 22.85$\pm$0.09 & 23.65$\pm$0.25 & 22.58$\pm$0.18 & 23.69$\pm$0.27\\
95 & 12:28:16.04 +44:06:38.03 & 22.69$\pm$0.09 & 22.61$\pm$0.08 & 22.61$\pm$0.08 & 22.65$\pm$0.08 & 22.5$\pm$0.08 & 20.35$\pm$0.1 & 22.71$\pm$0.08 & 23.5$\pm$0.14 & 22.58$\pm$0.3 & 23.94$\pm$0.4\\
96 & 12:28:15.92 +44:06:39.20 & 23.39$\pm$0.09 & 23.28$\pm$0.09 & 23.27$\pm$0.08 & 23.34$\pm$0.09 & 23.22$\pm$0.09 & 21.01$\pm$0.1 & 23.48$\pm$0.1 & 24.26$\pm$0.27 & 22.77$\pm$0.36 & 25.09$\pm$0.56\\
97 & 12:28:14.97 +44:06:46.12 & 25.93$\pm$0.33 & 25.08$\pm$0.15 & 24.69$\pm$0.13 & 24.01$\pm$0.11 & 24.12$\pm$0.12 & 21.91$\pm$0.21 & 23.26$\pm$0.1 & 23.29$\pm$0.15 & 22.26$\pm$0.18 & 23.12$\pm$0.13\\
98 & 12:28:11.52 +44:06:47.72 & 21.84$\pm$0.08 & 21.94$\pm$0.08 & 22.04$\pm$0.08 & 22.43$\pm$0.08 & 22.07$\pm$0.08 & 20.18$\pm$0.1 & 22.74$\pm$0.09 & 23.84$\pm$0.46 & 22.79$\pm$0.22 & 24.58$\pm$0.64\\
99 & 12:28:13.96 +44:06:49.28 & 20.95$\pm$0.09 & 20.99$\pm$0.08 & 21.16$\pm$0.09 & 21.62$\pm$0.09 & 20.57$\pm$0.08 & 18.42$\pm$0.09 & 21.62$\pm$0.09 & 22.3$\pm$0.13 & 20.95$\pm$0.15 & 22.62$\pm$0.14
\enddata
\tablenotetext{}{(1) The identification number of the source.}
\tablenotetext{}{(2) Right Ascension  and Declination in J2000 coordinates.}
\tablenotetext{}{(3-12) AB magnitude of each source in the corresponding filter as determined by performing aperture photometry.}
\end{deluxetable}
\end{longrotatetable}

%%%%%%%%%%%%%%%%%%%%%%%%%%%%%%%%%%%%%%%%%%%%%%%%%%%%%%

\startlongtable
\begin{deluxetable}{lcrrr}
\centering 
\tablecolumns{5}
%%\tabletypesize{\tiny}
\tablewidth{0pt}
\tablecaption{Source Properties}
\tablehead{\colhead{ID} & \colhead{Age} & \colhead{E(B-V)} & \colhead{Log(Mass)} & \colhead{$\chi^{2}_{red}$}\\\colhead{ } & \colhead{Myr} & \colhead{mag} & \colhead{M$_{\odot}$}}
\label{tab:Source Properties Table}
\startdata
1 & $5^{+0}_{-0}$ & $0.0^{+0.04}_{-0.0}$ & $2.54^{+0.057}_{-0.0}$ & 3.74\\
2 & $8^{+6}_{-1}$ & $0.46^{+0.16}_{-0.14}$ & $2.99^{+0.514}_{-0.277}$ & 8.26\\
3 & $3^{+1}_{-0}$ & $0.46^{+0.1}_{-0.08}$ & $3.12^{+0.223}_{-0.098}$ & 5.13\\
4 & $5^{+0}_{-0}$ & $0.0^{+0.04}_{-0.0}$ & $2.8^{+0.055}_{-0.0}$ & 15.41\\
5 & $9000^{+0}_{-8994}$ & $0.04^{+0.92}_{-0.04}$ & $4.53^{+0.085}_{-1.909}$ & 2.44\\
6 & $4^{+1}_{-0}$ & $0.8^{+0.24}_{-0.26}$ & $2.92^{+0.238}_{-0.192}$ & 8.92\\
7 & $5^{+0}_{-1}$ & $0.0^{+0.08}_{-0.0}$ & $2.86^{+0.119}_{-0.082}$ & 7.64\\
8 & $9000^{+0}_{-8994}$ & $0.0^{+0.96}_{-0.0}$ & $4.81^{+0.097}_{-1.949}$ & 5.9\\
9 & $7^{+5}_{-2}$ & $0.2^{+0.36}_{-0.14}$ & $2.5^{+0.638}_{-0.199}$ & 10.16\\
10 & $2000^{+1000}_{-1940}$ & $0.0^{+0.68}_{-0.0}$ & $4.16^{+0.192}_{-0.444}$ & 2.78\\
11 & $4^{+0}_{-0}$ & $0.0^{+0.08}_{-0.0}$ & $2.7^{+0.116}_{-0.0}$ & 4.45\\
12 & $9^{+31}_{-2}$ & $0.56^{+0.16}_{-0.14}$ & $3.56^{+0.554}_{-0.368}$ & 7.76\\
13 & $6^{+8994}_{-1}$ & $1.32^{+0.22}_{-0.92}$ & $3.41^{+1.58}_{-0.388}$ & 8.9\\
14 & $4^{+0}_{-1}$ & $0.3^{+0.08}_{-0.12}$ & $2.66^{+0.106}_{-0.264}$ & 5.56\\
15 & $5^{+9}_{-0}$ & $0.0^{+0.08}_{-0.0}$ & $2.48^{+0.434}_{-0.035}$ & 16.64\\
16 & $5^{+0}_{-0}$ & $0.0^{+0.08}_{-0.0}$ & $3.33^{+0.103}_{-0.0}$ & 10.59\\
17 & $6^{+8994}_{-1}$ & $0.78^{+0.22}_{-0.78}$ & $2.85^{+1.598}_{-0.395}$ & 1.86\\
18 & $4^{+0}_{-2}$ & $0.3^{+0.08}_{-0.12}$ & $3.52^{+0.107}_{-0.227}$ & 4.55\\
19 & $4^{+0}_{-0}$ & $0.72^{+0.14}_{-0.16}$ & $2.55^{+0.164}_{-0.187}$ & 1.23\\
20 & $9^{+11}_{-2}$ & $0.7^{+0.16}_{-0.14}$ & $3.18^{+0.368}_{-0.273}$ & 4.17\\
21 & $9^{+5}_{-2}$ & $0.32^{+0.1}_{-0.14}$ & $4.09^{+0.373}_{-0.384}$ & 6.8\\
22 & $7^{+3}_{-1}$ & $0.76^{+0.24}_{-0.14}$ & $3.29^{+0.514}_{-0.139}$ & 13.31\\
23 & $9^{+5}_{-2}$ & $0.26^{+0.12}_{-0.12}$ & $3.16^{+0.369}_{-0.35}$ & 15.6\\
24 & $9000^{+0}_{-7000}$ & $0.42^{+0.2}_{-0.12}$ & $5.39^{+0.082}_{-0.505}$ & 14.0\\
25 & $4^{+0}_{-0}$ & $0.18^{+0.08}_{-0.08}$ & $3.37^{+0.114}_{-0.114}$ & 2.55\\
26 & $5^{+0}_{-1}$ & $0.8^{+0.1}_{-0.12}$ & $3.69^{+0.1}_{-0.165}$ & 13.0\\
27 & $5^{+0}_{-1}$ & $0.1^{+0.08}_{-0.08}$ & $2.6^{+0.114}_{-0.114}$ & 2.56\\
28 & $6^{+0}_{-0}$ & $0.16^{+0.08}_{-0.08}$ & $3.28^{+0.105}_{-0.105}$ & 9.82\\
29 & $6^{+8994}_{-0}$ & $1.2^{+0.14}_{-0.98}$ & $4.09^{+1.615}_{-0.444}$ & 11.01\\
30 & $5^{+0}_{-0}$ & $0.5^{+0.06}_{-0.08}$ & $4.41^{+0.072}_{-0.096}$ & 19.61\\
31 & $6^{+0}_{-1}$ & $0.0^{+0.08}_{-0.0}$ & $2.46^{+0.103}_{-0.05}$ & 5.12\\
32 & $1^{+0}_{-0}$ & $0.3^{+0.08}_{-0.08}$ & $3.45^{+0.101}_{-0.101}$ & 12.27\\
33 & $7^{+8993}_{-0}$ & $0.98^{+0.1}_{-0.72}$ & $3.54^{+1.9}_{-0.067}$ & 3.64\\
34 & $20^{+20}_{-12}$ & $0.46^{+0.16}_{-0.1}$ & $4.38^{+0.216}_{-0.544}$ & 5.57\\
35 & $5^{+0}_{-1}$ & $0.12^{+0.08}_{-0.1}$ & $3.0^{+0.112}_{-0.2}$ & 3.88\\
36 & $5^{+1}_{-0}$ & $0.04^{+0.1}_{-0.04}$ & $3.16^{+0.151}_{-0.085}$ & 6.34\\
37 & $5^{+0}_{-0}$ & $0.0^{+0.0}_{-0.0}$ & $3.21^{+0.0}_{-0.0}$ & 40.32\\
38 & $9^{+11}_{-2}$ & $0.36^{+0.12}_{-0.12}$ & $3.52^{+0.372}_{-0.355}$ & 24.66\\
39 & $6^{+9}_{-0}$ & $1.16^{+0.2}_{-0.48}$ & $3.34^{+0.247}_{-0.418}$ & 6.16\\
40 & $4^{+1}_{-0}$ & $0.16^{+0.06}_{-0.08}$ & $3.75^{+0.159}_{-0.099}$ & 14.67\\
41 & $6^{+3}_{-0}$ & $1.14^{+0.2}_{-0.44}$ & $4.02^{+0.099}_{-0.402}$ & 4.58\\
42 & $4^{+0}_{-1}$ & $0.06^{+0.08}_{-0.06}$ & $2.47^{+0.128}_{-0.184}$ & 2.18\\
43 & $9^{+21}_{-2}$ & $0.82^{+0.14}_{-0.14}$ & $3.27^{+0.458}_{-0.348}$ & 8.61\\
44 & $6^{+8994}_{-0}$ & $1.16^{+0.14}_{-0.98}$ & $4.08^{+1.589}_{-0.451}$ & 8.83\\
45 & $6^{+8994}_{-0}$ & $1.44^{+0.12}_{-0.94}$ & $3.64^{+1.598}_{-0.368}$ & 9.11\\
46 & $4^{+0}_{-0}$ & $0.08^{+0.08}_{-0.08}$ & $2.33^{+0.114}_{-0.114}$ & 2.85\\
47 & $5^{+0}_{-0}$ & $0.28^{+0.08}_{-0.08}$ & $3.06^{+0.111}_{-0.111}$ & 6.62\\
48 & $4^{+0}_{-0}$ & $0.06^{+0.08}_{-0.06}$ & $2.45^{+0.115}_{-0.087}$ & 5.0\\
49 & $7^{+8993}_{-1}$ & $0.88^{+0.4}_{-0.68}$ & $3.37^{+1.927}_{-0.098}$ & 8.59\\
50 & $6^{+74}_{-1}$ & $0.76^{+0.26}_{-0.32}$ & $2.97^{+0.793}_{-0.377}$ & 5.91\\
51 & $5^{+0}_{-0}$ & $0.32^{+0.08}_{-0.08}$ & $3.1^{+0.104}_{-0.104}$ & 5.99\\
52 & $5^{+0}_{-1}$ & $0.18^{+0.06}_{-0.08}$ & $3.26^{+0.079}_{-0.192}$ & 11.66\\
53 & $1^{+0}_{-0}$ & $0.28^{+0.08}_{-0.06}$ & $3.86^{+0.102}_{-0.076}$ & 16.44\\
54 & $3^{+1}_{-2}$ & $0.5^{+0.14}_{-0.14}$ & $2.23^{+0.288}_{-0.163}$ & 1.62\\
55 & $5^{+0}_{-0}$ & $0.06^{+0.08}_{-0.06}$ & $2.74^{+0.118}_{-0.089}$ & 7.95\\
56 & $4^{+0}_{-0}$ & $0.02^{+0.08}_{-0.02}$ & $1.96^{+0.128}_{-0.032}$ & 1.71\\
57 & $5^{+0}_{-0}$ & $0.14^{+0.08}_{-0.06}$ & $3.43^{+0.097}_{-0.073}$ & 17.82\\
58 & $4^{+0}_{-1}$ & $0.04^{+0.08}_{-0.04}$ & $2.45^{+0.104}_{-0.15}$ & 5.18\\
59 & $1^{+1}_{-0}$ & $0.38^{+0.1}_{-0.08}$ & $2.47^{+0.121}_{-0.097}$ & 7.3\\
60 & $5^{+0}_{-0}$ & $0.0^{+0.04}_{-0.0}$ & $2.17^{+0.062}_{-0.0}$ & 5.73\\
61 & $5^{+0}_{-0}$ & $0.0^{+0.04}_{-0.0}$ & $2.83^{+0.059}_{-0.0}$ & 7.04\\
62 & $5^{+0}_{-0}$ & $0.22^{+0.08}_{-0.06}$ & $2.89^{+0.105}_{-0.079}$ & 4.12\\
63 & $3^{+1}_{-2}$ & $0.46^{+0.24}_{-0.22}$ & $2.04^{+0.406}_{-0.218}$ & 0.35\\
64 & $5^{+0}_{-0}$ & $0.0^{+0.02}_{-0.0}$ & $3.35^{+0.03}_{-0.0}$ & 10.28\\
65 & $3^{+1}_{-2}$ & $0.56^{+0.18}_{-0.2}$ & $2.5^{+0.272}_{-0.164}$ & 2.32\\
66 & $1^{+2}_{-0}$ & $0.18^{+0.14}_{-0.1}$ & $2.4^{+0.146}_{-0.146}$ & 2.37\\
67 & $4^{+0}_{-1}$ & $0.12^{+0.08}_{-0.12}$ & $2.47^{+0.113}_{-0.272}$ & 3.57\\
68 & $5^{+0}_{-0}$ & $0.88^{+0.14}_{-0.12}$ & $3.32^{+0.134}_{-0.115}$ & 4.34\\
69 & $5^{+0}_{-0}$ & $0.0^{+0.06}_{-0.0}$ & $2.66^{+0.089}_{-0.0}$ & 2.71\\
70 & $5^{+0}_{-1}$ & $0.0^{+0.02}_{-0.0}$ & $2.96^{+0.03}_{-0.038}$ & 8.63\\
71 & $13^{+2987}_{-2}$ & $0.88^{+0.08}_{-0.54}$ & $3.66^{+1.009}_{-0.108}$ & 4.68\\
72 & $5^{+0}_{-1}$ & $0.32^{+0.1}_{-0.08}$ & $2.89^{+0.138}_{-0.184}$ & 4.79\\
73 & $3^{+0}_{-0}$ & $0.12^{+0.08}_{-0.08}$ & $2.6^{+0.12}_{-0.12}$ & 1.66\\
74 & $1^{+2}_{-0}$ & $0.12^{+0.1}_{-0.08}$ & $2.8^{+0.113}_{-0.124}$ & 1.78\\
75 & $4^{+0}_{-1}$ & $0.46^{+0.12}_{-0.2}$ & $2.46^{+0.166}_{-0.4}$ & 2.43\\
76 & $6^{+0}_{-1}$ & $0.0^{+0.24}_{-0.0}$ & $2.22^{+0.409}_{-0.0}$ & 7.28\\
77 & $3^{+1}_{-2}$ & $0.48^{+0.28}_{-0.18}$ & $2.17^{+0.481}_{-0.215}$ & 1.56\\
78 & $6^{+0}_{-0}$ & $0.0^{+0.06}_{-0.0}$ & $2.28^{+0.089}_{-0.0}$ & 4.85\\
79 & $2^{+1}_{-1}$ & $0.24^{+0.12}_{-0.12}$ & $2.62^{+0.155}_{-0.155}$ & 2.71\\
80 & $3^{+1}_{-2}$ & $0.58^{+0.2}_{-0.16}$ & $2.41^{+0.356}_{-0.177}$ & 2.67\\
81 & $20^{+20}_{-11}$ & $0.16^{+0.14}_{-0.14}$ & $4.26^{+0.164}_{-0.434}$ & 4.29\\
82 & $6^{+1}_{-1}$ & $1.1^{+0.34}_{-0.5}$ & $4.17^{+0.361}_{-0.47}$ & 2.99\\
83 & $4^{+1}_{-0}$ & $0.06^{+0.08}_{-0.06}$ & $2.6^{+0.207}_{-0.093}$ & 2.46\\
84 & $1^{+1}_{-0}$ & $0.44^{+0.1}_{-0.1}$ & $3.13^{+0.129}_{-0.129}$ & 4.16\\
85 & $5^{+0}_{-1}$ & $0.06^{+0.08}_{-0.06}$ & $2.49^{+0.112}_{-0.16}$ & 3.76\\
86 & $12^{+88}_{-4}$ & $0.26^{+0.08}_{-0.22}$ & $4.43^{+0.422}_{-0.506}$ & 1.37\\
87 & $4^{+0}_{-0}$ & $0.18^{+0.08}_{-0.08}$ & $2.42^{+0.114}_{-0.114}$ & 3.98\\
88 & $5^{+0}_{-0}$ & $0.0^{+0.08}_{-0.0}$ & $2.59^{+0.123}_{-0.0}$ & 1.11\\
89 & $1^{+1}_{-0}$ & $0.34^{+0.1}_{-0.08}$ & $2.49^{+0.113}_{-0.122}$ & 3.0\\
90 & $3^{+1}_{-2}$ & $0.14^{+0.08}_{-0.14}$ & $2.15^{+0.216}_{-0.151}$ & 2.63\\
91 & $4^{+0}_{-0}$ & $0.1^{+0.08}_{-0.08}$ & $2.45^{+0.12}_{-0.121}$ & 1.42\\
92 & $4^{+0}_{-0}$ & $0.06^{+0.1}_{-0.06}$ & $1.97^{+0.156}_{-0.094}$ & 2.33\\
93 & $6^{+0}_{-1}$ & $0.26^{+0.14}_{-0.08}$ & $3.26^{+0.202}_{-0.101}$ & 11.79\\
94 & $4^{+1}_{-0}$ & $0.14^{+0.08}_{-0.1}$ & $2.36^{+0.196}_{-0.146}$ & 2.54\\
95 & $4^{+1}_{-0}$ & $0.18^{+0.1}_{-0.08}$ & $2.43^{+0.184}_{-0.116}$ & 5.91\\
96 & $4^{+1}_{-0}$ & $0.2^{+0.08}_{-0.08}$ & $2.18^{+0.133}_{-0.122}$ & 3.63\\
97 & $5^{+0}_{-1}$ & $0.88^{+0.14}_{-0.14}$ & $2.81^{+0.141}_{-0.141}$ & 2.44\\
98 & $4^{+1}_{-0}$ & $0.02^{+0.08}_{-0.02}$ & $2.36^{+0.194}_{-0.03}$ & 2.75\\
99 & $2^{+2}_{-1}$ & $0.0^{+0.14}_{-0.0}$ & $2.74^{+0.223}_{-0.059}$ & 6.88
\enddata
\tablenotetext{}{ID number, age, color excess, mass, and reduced $\chi^{2}$ obtained from SED fitting.}
\end{deluxetable}

\end{document}